\documentclass[letter]{article}

\usepackage[total={6in,8in}]{geometry}
\usepackage[table]{xcolor}

\usepackage{amsmath}
\usepackage{amssymb}
\usepackage{url}
\usepackage{epsfig}
\usepackage{graphicx}
\usepackage{algorithm}
\usepackage{algorithmicx}
\usepackage[noend]{algpseudocode}
\usepackage{tikz}
\usepackage{balance}
\usepackage{hyperref}

\usepackage{rotating}
\usepackage{subfig}
\usepackage{color}
\usepackage{multirow}
\usepackage{amsmath}
\usepackage{xspace}
\usepackage{bm}

\usepackage{enumitem}
\usepackage{lipsum}
\usepackage[font=footnotesize,labelfont=bf]{caption}
\let\OLDthebibliography\thebibliography
\renewcommand\thebibliography[1]{
  \OLDthebibliography{#1}
  \setlength{\parskip}{0pt}
  \setlength{\itemsep}{0pt plus 0.3ex}
}

\newcommand\smallerfont{\fontsize{9}{10}\selectfont}

\newcommand{\myspace}[1]{} 
\newcommand{\spgemm}{{\sc s}p{\sc gemm}}

\newcommand\gpu{{\sc gpu}}
\newcommand\knl{{\sc knl}}

\newcommand\ddr{{\sc ddr}}
\newcommand\cpu{{\sc cpu}}
\newcommand\hbm{{\sc hbm}}
\newcommand\uvm{{\sc uvm}}

\def\cca#1{\cellcolor{black!#10}\ifnum #1>5\color{white}\fi{#1}}

\newcommand{\avgdeg}[1]{$\delta_{#1}$}

\newcommand\imscale {0.5}
\newcommand\imtscale {0.5}
\newcommand\tabscale {0.75}
\begin{document}

\title{\LARGE \bf Sparse Matrix-Matrix Multiplication on Multilevel Memory Architectures : Algorithms and Experiments}

\author{
Mehmet Deveci, Simon D. Hammond, \\Michael M. Wolf, and Sivasankaran Rajamanickam,\\ 
\{mndevec, sdhammo, mmwolf, srajama\}@sandia.gov \\ Sandia National Laboratories, Albuquerque, NM \\ SAND2018-3428 R}

\maketitle

\date{}



\begin{abstract}
\noindent
Architectures with multiple classes of memory media are becoming a
common part of mainstream supercomputer deployments. So called
multi-level memories offer differing characteristics for each memory
component including variation in bandwidth, latency and capacity. 
This paper investigates the performance of sparse matrix
multiplication kernels on two leading high-performance
computing architectures -- Intel's Knights Landing processor
and NVIDIA's Pascal GPU.
We describe a data placement method and a chunking-based algorithm 
for our kernels that exploits the existence of the multiple memory spaces 
in each hardware platform. We evaluate the performance of these
methods w.r.t. standard algorithms using the auto-caching mechanisms.
Our results show that standard algorithms that exploit cache reuse
performed as well as multi-memory-aware algorithms for architectures 
such as \knl{}s where the memory subsystems have similar latencies.
However, for architectures such as \gpu{}s where memory subsystems differ
significantly in both bandwidth and latency, multi-memory-aware methods
are crucial for good performance. 
In addition, our new approaches permit the user to run problems that require larger 
capacities than the fastest memory of each compute node without depending 
on the software-managed cache mechanisms.
\end{abstract}

\section{Introduction}
\label{sec:intro}

\noindent
Complex memory subsystems with multiple levels of memory are part of
both recently deployed supercomputers \cite{trinityweb,antypas2014cori,
thetaweb} and proposed future
deployments~\cite{neely2017application}. The use of different types of
memory is driven by multiple factors including cost,
performance and energy. This results in significant variation in memory bandwidth
and latency. 
Additionally, with growing diversity of solutions 
due to the inclusion of byte-addressable non-volatile memories,
this discrepancy in memory bandwidth and latency is expected to increase.
Such systems will experience much greater variability in memory access times due to
the asymmetry of read/write access times.
These complexities lead to the question ``do algorithms have 
to be redesigned to account for these multi-memory subspaces?''.

Bender {\em et al.}. considered two-level memory and corroborated the need for
multilevel algorithms using simulation~\cite{bender2015two}.
That particular study focused on sorting algorithms and was limited to
simulation due to the unavailability of hardware. It used Sandia National
Laboratories' Structural Simulation Toolkit (SST)
\cite{rodrigues2011structural} simulator for projecting the
behavior of multi-level memory aware algorithms. This study was extended
later to a $k$-means clustering in~\cite{bender2015k}.  In this paper,
we consider this theoretical work in the context of a new linear algebra kernel,
and evaluate it on two different hardware with complex memory subsystems
and varying levels of concurrency. 

Our primary focus is on the development and optimization of sparse
matrix-matrix multiplication (\spgemm{}) kernels for complex memory
subsystems. This kernel is of interest in a variety scientific computing
applications as \spgemm{} is the most expensive kernel in many
algorithms including the setup phase of multigrid methods. Sparse
matrix-matrix multiplication also has applicability in data analysis
problems as it is a foundational kernel for the GraphBLAS
effort~\cite{mattson2013standards}. Many graph analysis problems can be
expressed in terms of \spgemm{}
\cite{bulucc2011combinatorial,wolf2017fast}. Hence optimizing \spgemm{}
on multilevel memory architectures has the potential to impact a wide
variety of applications. The baseline \spgemm{} algorithm used in
this study has been shown to outperform equivalent vendor 
implementations~\cite{deveci2017performance, deveci2018multi} 
(On average $12\%$ and $2.56\times$ faster than best MKL method 
and cuSPARSE~\cite{deveci2018multi}).
This algorithm has become the default in Trilinos~\cite{heroux2005overview}. 
Hence, improving the performance of this \spgemm{} algorithm will directly
impact a number of exascale computing applications that rely on
Trilinos.  This baseline algorithm was also used as a kernel within the
linear-algebra based triangle counting method~\cite{wolf2017fast} where
Wolf {\em et al.}  demonstrated that linear-algebra based methods can
be as fast as graph traversal based methods when using the \spgemm{}
algorithm.



Since the Bender {\em et al.} paper, Intel has provided the option to
treat the high bandwidth memory (\hbm{}) as a ``cache'' on \knl{}s. This is despite the latency costs of \hbm{}
being higher than one would typically expect from a ``cache''. Software managed
mode is still an option for applications that will not fit into \hbm{}, but where
cache mode does not perform well. On the \gpu{}, memory accesses to host
pinned \ddr{} memory is allowed and Unified Memory (\uvm{}) accesses work similar to
the ``cache'' mode.  All these options allow a number of variations of the
standard algorithms that can be run in real hardware. This raises the question
``When and where are two-level algorithms needed for an optimized kernel like
\spgemm{}?''. We develop a two-level scratchpad memory-aware
``chunked algorithm'', and compare it to a highly optimized
traditional, one-level aware algorithm with all the vendor provided options for
ease-of-use of multilevel memories.

The contributions of this paper are summarized below.
\begin{itemize}
\item We provide a thorough performance analysis of the baseline state-of-art 
\spgemm{} method using different memory subsystems on \gpu{}s and \knl{}s. This 
analysis studies the effects of access patterns, selective data placement methods 
and cache-modes provided by the architectures on the performance of \spgemm{}. 

\item We identify which data structures are critical in the performance of \spgemm{} on
these architectures. Using this, we propose a data placement method and chunking-based 
algorithm that better exploits multi-memory subsystems. 

\item We evaluate the performance of these methods w.r.t. standard algorithms on \gpu{} and 
\knl{}s. Using the different characteristics of the memory subsystems in these architectures, 
we demonstrate where and when the proposed chunked algorithm and data placement strategies 
are useful for the \spgemm{} kernel.
\end{itemize}




\myspace{-1.5ex}
\section{Background}
\label{sec:background}
\myspace{-0.5ex}

\noindent
Some of the earliest work on the porting of algorithms to multiple pools
of memory with differing performance occured during the very earliest
days of computing. In previous machines, application working sets would
have been written to physical media and then loaded back into the
limited volatile memory stores for processing in blocks.

Recently the emergence of GPU-based
accelerators~\cite{lindholm2008nvidia} has reignited interest in the
algorithms field due to the limited memory capacities available on GPU-based
compute engines. Our anecdotal experience has
been that many application developers porting to GPU-based systems have
opted to house their data sets entirely resident in the GPU memory
because the cost of transfer has been too high to
sufficiently amortize during execution. The recent addition of NVIDIA's
NVLINK bus~\cite{foley2017ultra}, which provides much higher data
transfer bandwidth, has the potential to change this balance in favor of
buffered algorithms.

The arrival of Intel's Knights Landing processor~\cite{jeffers2016intel,
sodani2015knights}, which was one of the first modern HPC processors to
provide multiple classes of memory on a single die, provided new
opportunities for algorithmic optimization because data structures no
longer needed to be entirely resident in any specific class of memory.
Instead, developers have been given additional choices, where data
structure {\em placement} can signficiantly affect
performance~\cite{voskuilen2016asc, voskuilenevaluating, detar2016milc,
li2017exploring}. 

Intel's forthcoming Optane class memory~\cite{foong2016storage,
wu2017early}, will add additional algorithm challenges and
opportunities. In these systems, byte-addressable non-volatile memory
technology will be available, enabling much larger memory capacities 
than ever before. However, this increased capacity will
sacrifice performance and present significant asymmetry in read and
write operation times. The use of multi-memory aware methods and a development focus on
data reuse has the potential to alleviate some of these concerns but
will require new families of algorithms to be developed.

Bender {\em et al.}~\cite{bender2015k} proposed a theoretical model 
that describes four necessary properties for
algorithms to benefit from a chunked variant: (1) memory boundedness;
(2) computation that can be broken down in scratch pad size chunks; (3)
computation where cache chunking is insufficient, and, (4) computation
that can reuse data loaded into a scratchpad. This follows from their
earlier work on sorting~\cite{bender2015two} which showed that 
chunking based algorithms can benefit sorting. Although this work
provides a thorough study, the proposed model does not consider the 
latency-based differences between the memory subsystems. Our work 
experimentally validates these properties on \knl{}s, and extends 
these studies for the memory subsystems with differing latency 
characterics, e.g. \gpu{}s.

There are multilevel
algorithms in other contexts such as between {\sc SSD} and \ddr{}. Zheng
{\em et al.}~\cite{Zheng2017SemiExternalMS} have shown that two level memory
algorithms are useful for sparse matrix times a dense vector/matrix
operation. This work is similar to our data placement strategy where
they hold the dense vector/matrix in memory. A recent work~\cite{joefficient}
can be considered similar to our chunking algorithm for
two level {\sc SSD} and \ddr{} memory algorithm for a sequential
implementation. 
The similarity in sequential chunking for a different level of memory 
subsystem \cite{joefficient} and our performance-portable approach 
validates the usefulness of the chunking strategy.

\subsection{Baseline {\sc SpGEMM} Method: {\sc kkmem}}

\noindent
In a recent work, we introduced a
\spgemm{} method, {\sc kkmem}~\cite{deveci2017performance, deveci2018multi},
designed to provide portability and cross-platform performance,
which performs well on various architectures such
as \cpu{}s, \knl{}s and \gpu{}s. {\sc kkmem} is a hierarchical, multithreaded 1D/2D row-wise algorithm.
It assigns the multiplication of each row to threads.
Different multiplications within the row are assigned to different
vector lanes. {\sc kkmem} is a two-phase algorithm: the number of
non-zeros in each row of the result matrix is calculated in the first
(symbolic) phase, then the actual values are computed in the
second (numeric) phase. 
{\em We focus on the numeric phase in this paper.}

{\sc kkmem} uses a compression technique to encode multiple columns of
the right hand side matrix with fewer integers. This reduces the number
of operations and the memory requirements in the symbolic phase
(See~\cite{deveci2017performance,wolf2017fast} for effects of
compression on different matrices). This also permits the use of bitwise
operations from union/intersection of different rows. 
Our implementation uses sparse hashmap-based accumulators together 
with a uniform memory pool for better memory scalability.

\begin{figure}
\begin{center}
\includegraphics[width=\imscale\columnwidth]{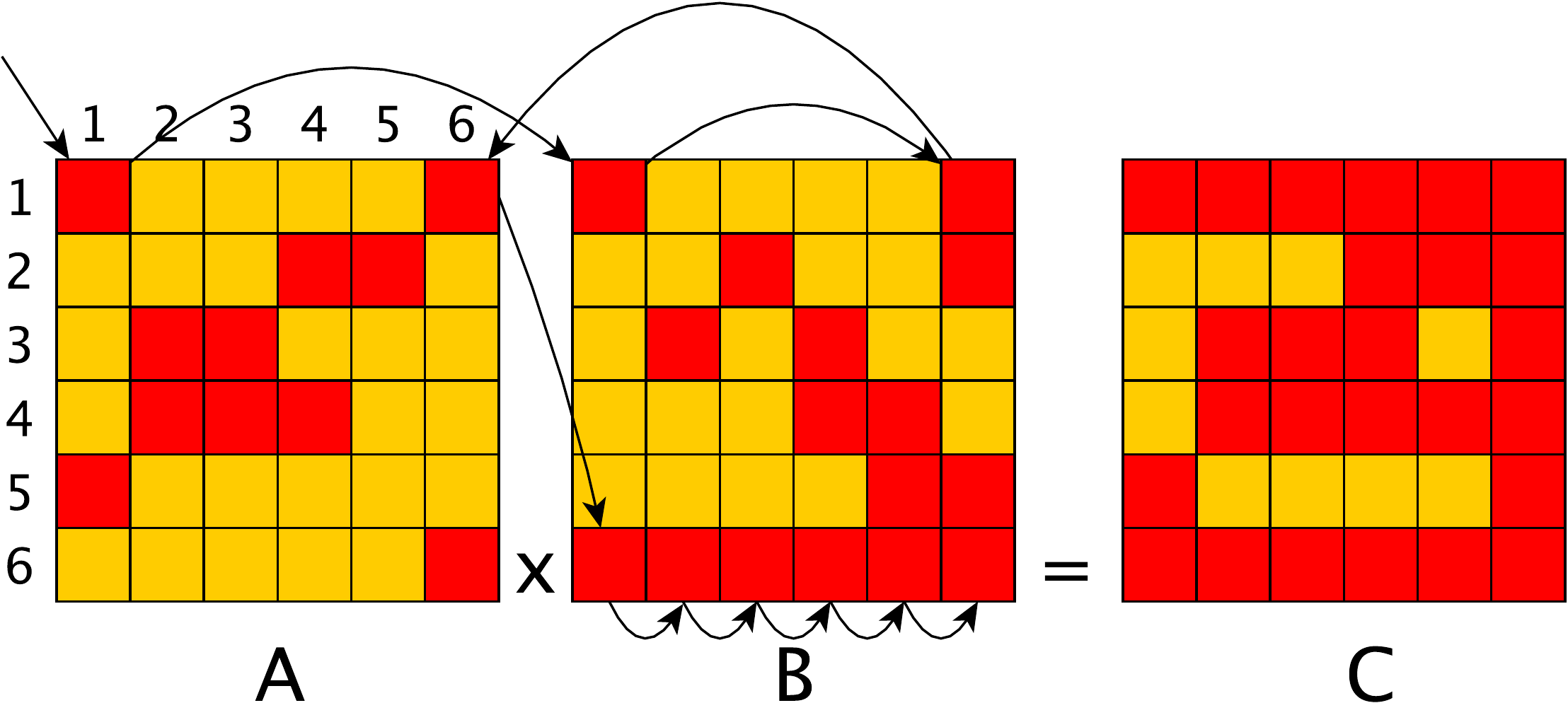}\label{fig:knlmcdram}
\end{center}
\caption{Example of \spgemm{} memory accesses. Arrows show the order of the read accesses for 
nonzero entries in $A$ and $B$.}
\label{fig:spgemmexp}
\end{figure}

Figure~\ref{fig:spgemmexp} gives an example of a simple multiplication.
Assume that a single thread performs the whole multiplication.
The rows of $A$ will be processed sequentially with each row being
multiplied with the nonzeroes of the corresponding rows of $B$, to compute a single row of
$C$. The arrows in the figure show the data access order for the
multiplication of the first row. The entry in column, $a_{11}$, of $A$ is read,
and multiplied with the entries in corresponding row of $B$ (first row).
Every non-zero in the first row of $B$ is read and multiplied and the
results are inserted into an accumulator. Then next column of $A$ ($a_{16}$) is read, and
algorithm reads through the $6th$ row of $B$. The results of the
multiplication are inserted into hashmap accumulator. When all multiplications
are completed the accumulator writes the values back to the corresponding 
row of $C$.

This \spgemm{} algorithm is a more challenging kernel than sorting for multilevel chunked
algorithms as it has been heavily optimized for efficient memory
accesses and cache reuse both on \gpu{}s and \knl{}
systems~\cite{deveci2017performance}. Bender {\em et al.} observed that
if cache chunking is sufficient one might not benefit from scratchpad
based algorithms. While the \spgemm{} implementation does not perform
cache chunking explicitly, it does optimize for spatial locality. There
are number of other variants \spgemm{} based on implementation on
different architectures, data structures used and partitioning employed.
Deveci {\em et al.}~\cite{deveci2017performance, deveci2018multi} provide a summary of
the most recent work that differ based on these parameters. Theoretical
approaches for avoiding communication in parallel \spgemm{} have been
extensively studied~\cite{akbudak2017exploiting, ballard2016hypergraph}. 
However, all these
approaches are limited to one level memory. Deveci {\em et al.} is also
one of the fastest implementations that we are aware of on \knl{}s and
\gpu{}s.
{Related benchmarks can be found in~\cite{deveci2017performance, deveci2018multi}, 
and {\url {https://github.com/kokkos/kokkos-kernels/wiki/SpGEMM_Benchmarks}}}. 
We adapt this highly scalable, performance-portable 
\spgemm{} algorithm to two-level memories on two different
architectures. 
This \spgemm{} algorithm has also been used as part of a fast
triangle counting approach on matrices from social networks
\cite{wolf2017fast}, which has been shown to be faster than highly optimized 
graph-based implementations. Using this, we also demonstrate the effect of 
memory systems on triangle counting problem.

\section{Performance Analysis of {\sc kkmem}}
\label{sec:algo}

\noindent
We analyze the memory accesses of {\sc kkmem}~\cite{deveci2017performance} analytically
and experimentally on multilevel memories on \knl{}s and \gpu{}s.
These analyses establish the need for strategies for handling
multilevel memories differently on different architectures.

\subsection {{\sc kkmem} Access Patterns}
Based on the access patterns given in Figure~\ref{fig:spgemmexp} and explained in 
previous section, 
some key observations on these access patterns follow:

\begin{itemize} 

\item $A$ is read in a stream-like fashion. Each entry is read once and used in
many multiplications. The accesses to $A$ are regular regardless of its structure.
 $C$ is written in the same order as $A$. Each entry in $C$ is written only once
in a streamed fashion, and oblivious to its structure.
Single row $j$ of $B$ is accessed as many times as the non-zeros
in the column $j$ of $A$. If each column of $A$ has uniform degree (\avgdeg{}) we will
read $B$ \avgdeg{} times. These accesses to $B$ can be irregular based on the 
structure of $A$. 

\item The number of insertions to the accumulators are as many as the number of 
multiplications. Depending on the type of accumulator, $B$'s structure might 
affect the memory accesses. For example, 
the first row of $B$ would insert first and last positions of the dense accumulators,
resulting in non-localized memory accesses breaking {\bf spatial locality}. 
Having narrow $B$ rows (low bandwidth) improves the spatial locality 
for dense accumulators. On the other hand, accesses to sparse accumulators 
have high locality regardless of $B$'s column indices, since they use much
smaller memory.
This improves the locality in the hashmap accumulators. 
However, the structure of B might affect the number of 
hash comparisons based on the collusions when sparse hashmap accumulators are used
as in {\sc kkmem}.

\item {\bf Temporal locality} is exploited by accessing a recently accessed row of $B$. 
For example, the multiplication of first row requires the first and last rows of $B$. 
They are not accessed again until the multiplication of the last two rows of $A$. 
On the other hand,
multiplication of the third row of $A$ accesses the second and third rows of $B$. These rows 
are immediately used in the multiplication of the fourth row of $A$, and they are 
likely to be in the cache. As a result, having overlapping columns in consecutive rows 
of $A$ improves temporal locality.

\item {\bf Cache Prefetching} reduces the latency cost by prefetching the data based on a
heuristic which is typically spatial locality. When rows of $B$ are very sparse, prefetching might bring
data from the next row, which may never be used. 
For example, in Figure~\ref{fig:spgemmexp} multiplication of the first row of $A$ might 
prefetch data from the second row of $B$ after accessing
first row of $B$. This row is not used in the multiplication of first row, and may not be used in the 
following multiplications based on the structure of $A$.  
When the rows of $B$ gets larger, these dense rows are likely to be prefetched 
and useful flops can be achieved with a lower latency cost.

\end{itemize}

\noindent
In summary, we expect most memory cost of {\sc kkmem} to be based on
$B$ accesses depending on the structure of $A$. Denser rows in $B$ help the performance with
prefetching. Better performance is expected when consecutive rows of $A$
have similar columns (increasing temporal locality). The temporal
locality can be further improved using hypergraph/graph partitioning
methods~\cite{akbudak2017exploiting}.
However, we avoid such preprocessing as they
require large number of multiplications to amortize their cost.
Based on this analysis, it is not
clear whether the insufficient cache chunking property required by the
analysis of Bender {\em et al.} applies immediately to \spgemm{}. 
In addition, how the performance of the algorithm reacts to the 
bandwidth and latency differencies in the memory subsystems requires 
further analysis. These result are dependent on problem structure and 
architecture, which are studied below.

\subsection{Analysis on KNL}

\begin{figure}
\begin{center}

\subfloat[LaplaceR]{\includegraphics[height=0.20\columnwidth, width=0.60\columnwidth]{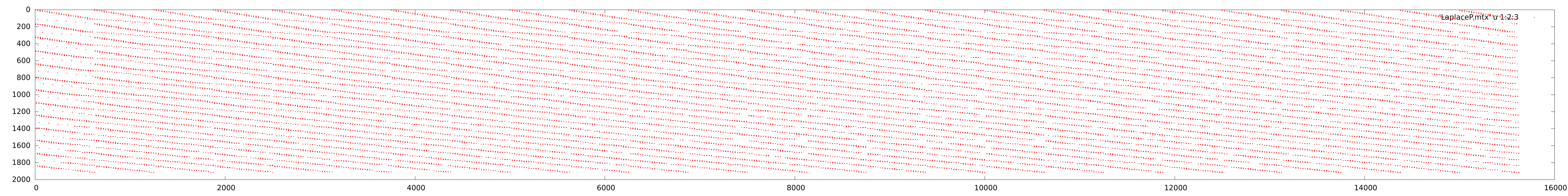}\label{fig:lapr}}
\subfloat[LaplaceA]{\includegraphics[width=0.40\columnwidth]{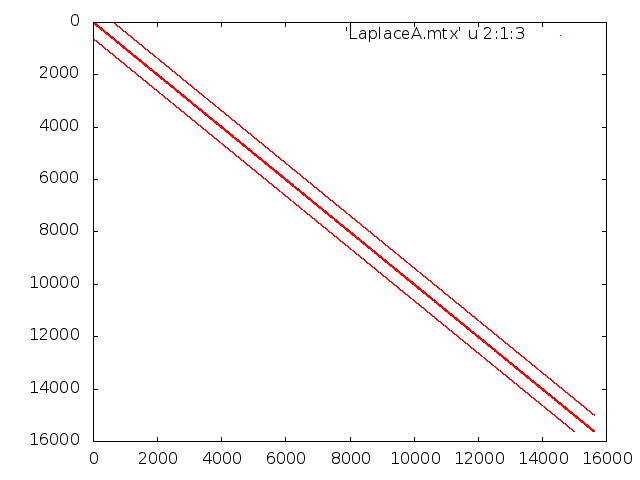}\label{fig:lapa}}



\end{center}
\caption{Examples of the Laplace matrices used in the paper}
\label{fig:lapmat}
\end{figure}

\noindent
We evaluate $48$ different multiplications from four
multigrid problem domains that are representative of different (but
typical) applications in the scientific computing domain at varying
scales. Triple products in the form of $A_c = R \times A_f \times P$ are a
key kernel in the setup of multigrid methods to generate a coarser
matrix $A_c$ from a fine matrix $A_f$. Figure~\ref{fig:lapmat} gives the
nonzero patterns for the $R$ and $A$ matrices for one of the problems --
Laplace3D. $P$ is transpose of $R$ in our examples. In these
scientific computing problems, $A$ matrices usually exhibit a
regular structure. The number of nonzeros per row of $A$ are $7$, $13$,
$27$, and $81$ on Laplace3D, BigStar2D, Brick3D, and Elasticity,
respectively. We expect a high temporal locality whenever
$A$ is on the left hand side. On the other hand, $R$ is a short and wide
rectangular matrix as shown in the figure. The rows have strided
columns, and consecutive rows do not have similar structure. Therefore,
we expect very low temporal and spatial locality when $R$ is on the
left hand side.

\begin{figure*}
\begin{center}
\subfloat[Laplace]{\includegraphics[width=\imscale\columnwidth]{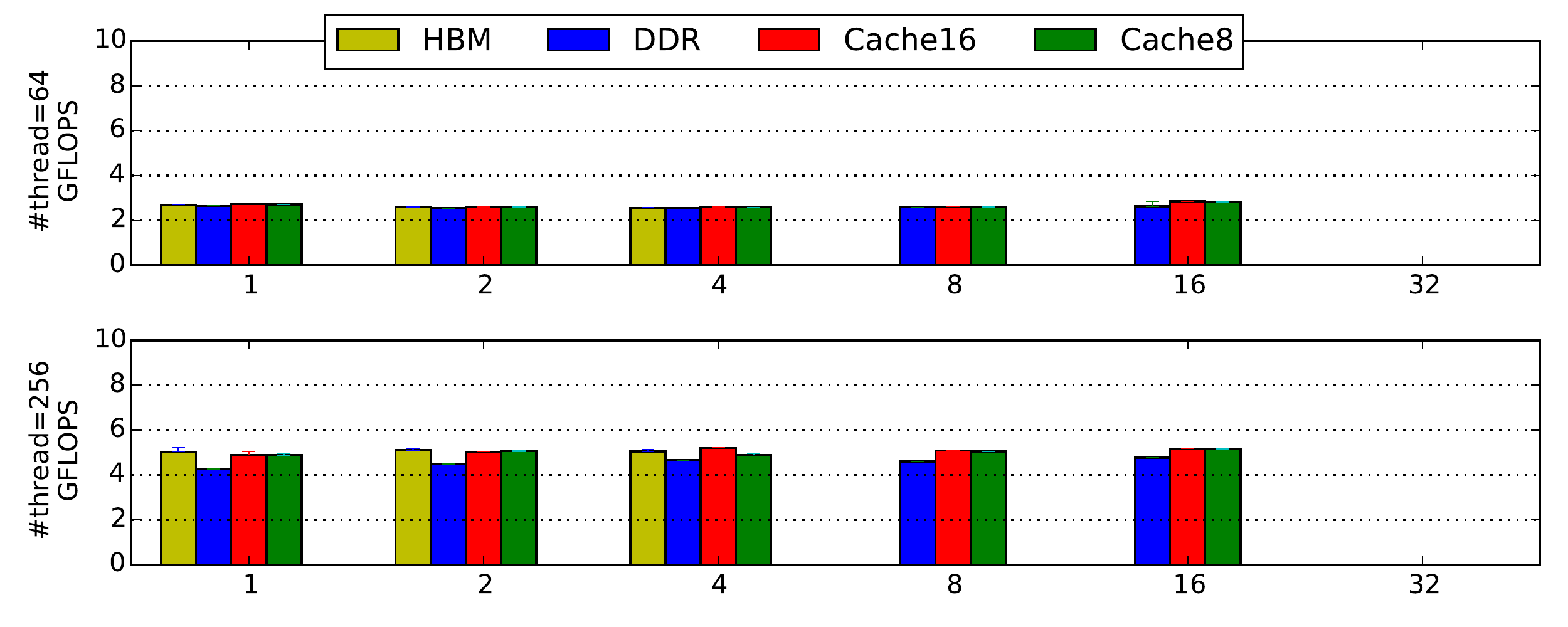}\label{fig:lapap}}
\subfloat[BigStar]{\includegraphics[width=\imscale\columnwidth]{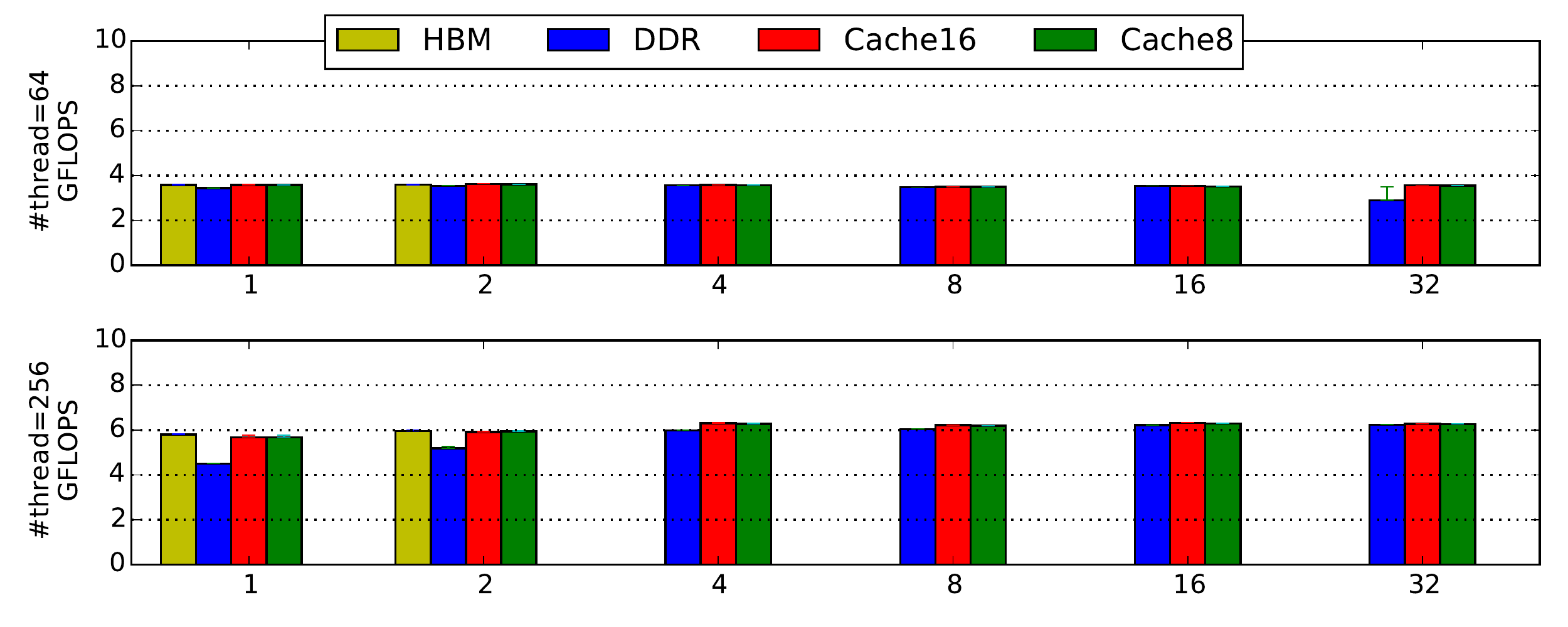}\label{fig:bigap}}

\subfloat[Brick]{\includegraphics[width=\imscale\columnwidth]{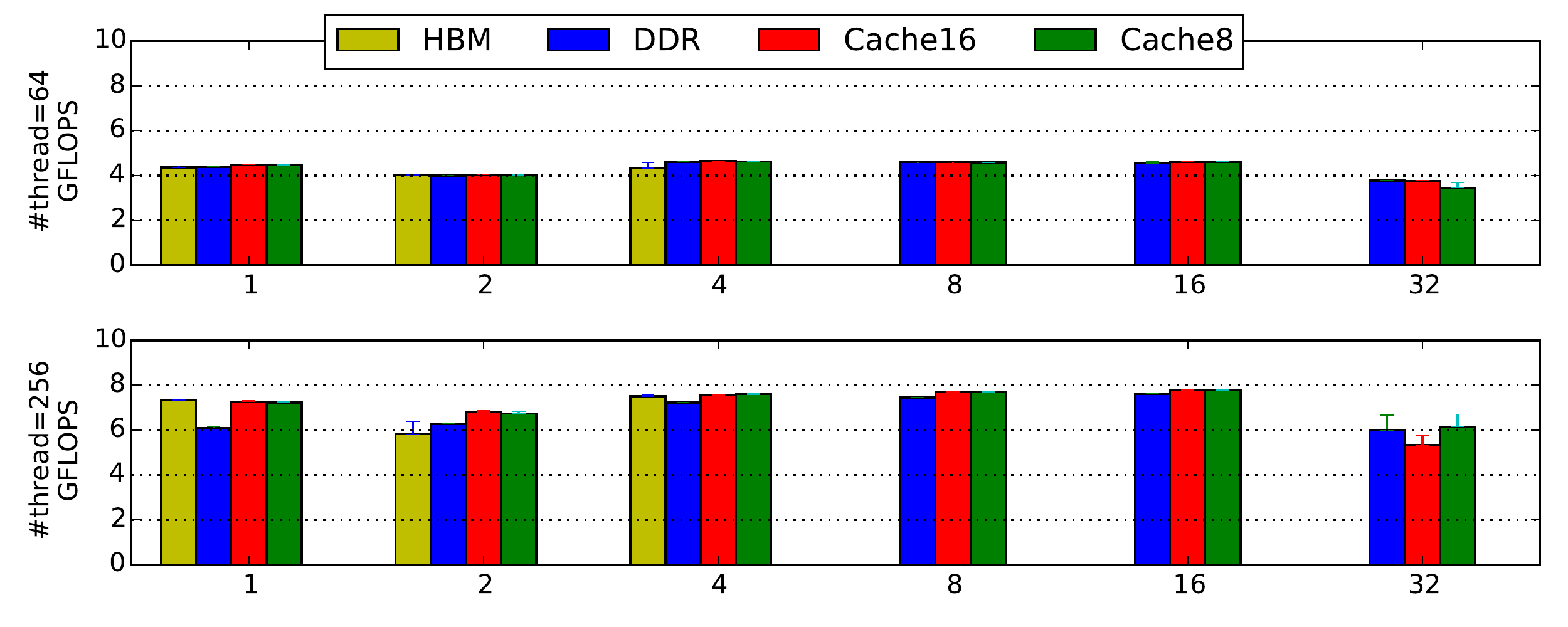}\label{fig:brickap}}
\subfloat[Elasticity]{\includegraphics[width=\imscale\columnwidth]{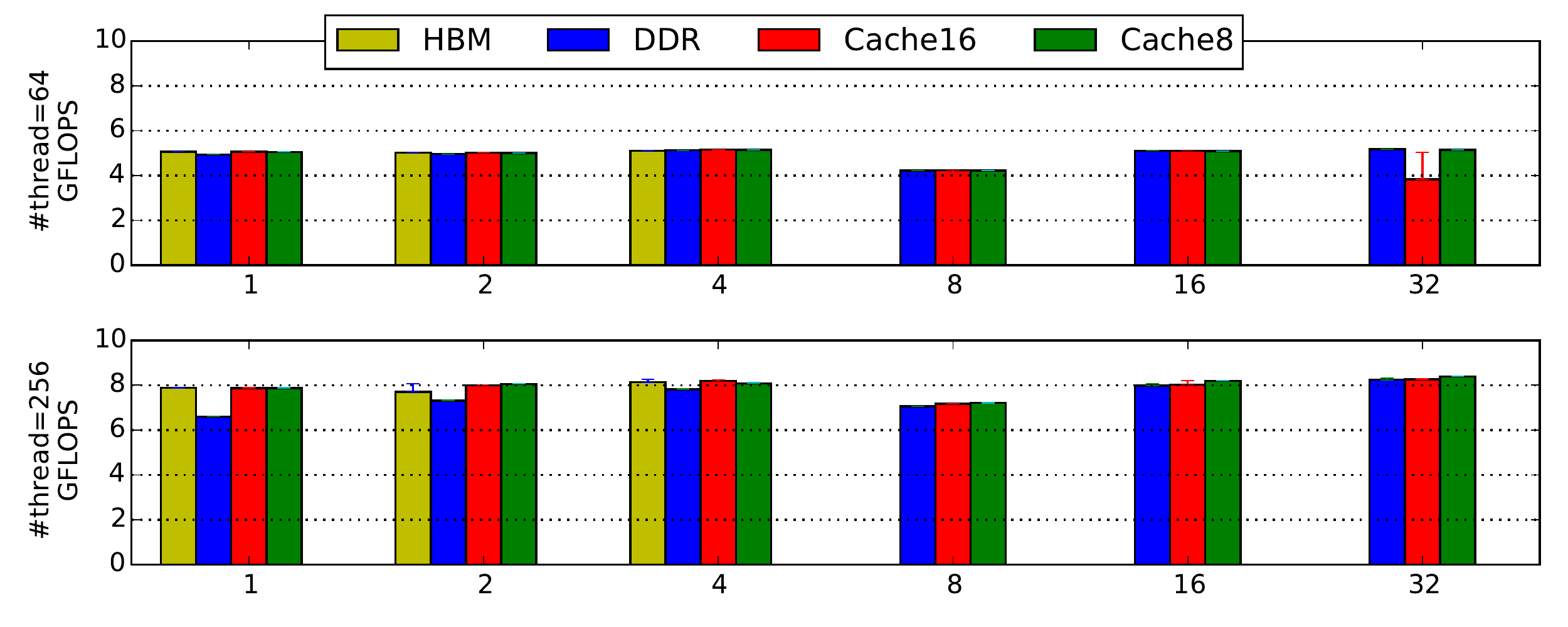}\label{fig:elasap}}

\end{center}
\caption{$A\times P$ GFLOPs on KNL. Top and bottom figures present the GFLOPs achieved on 64 and 256 threads, respectively. 
X axis refers to sizes of $A$ matrices (in GB) used in the experiments.}
\label{fig:knlap}
\end{figure*}

\begin{figure*}
\begin{center}

\subfloat[Laplace]{\includegraphics[width=\imscale\columnwidth]{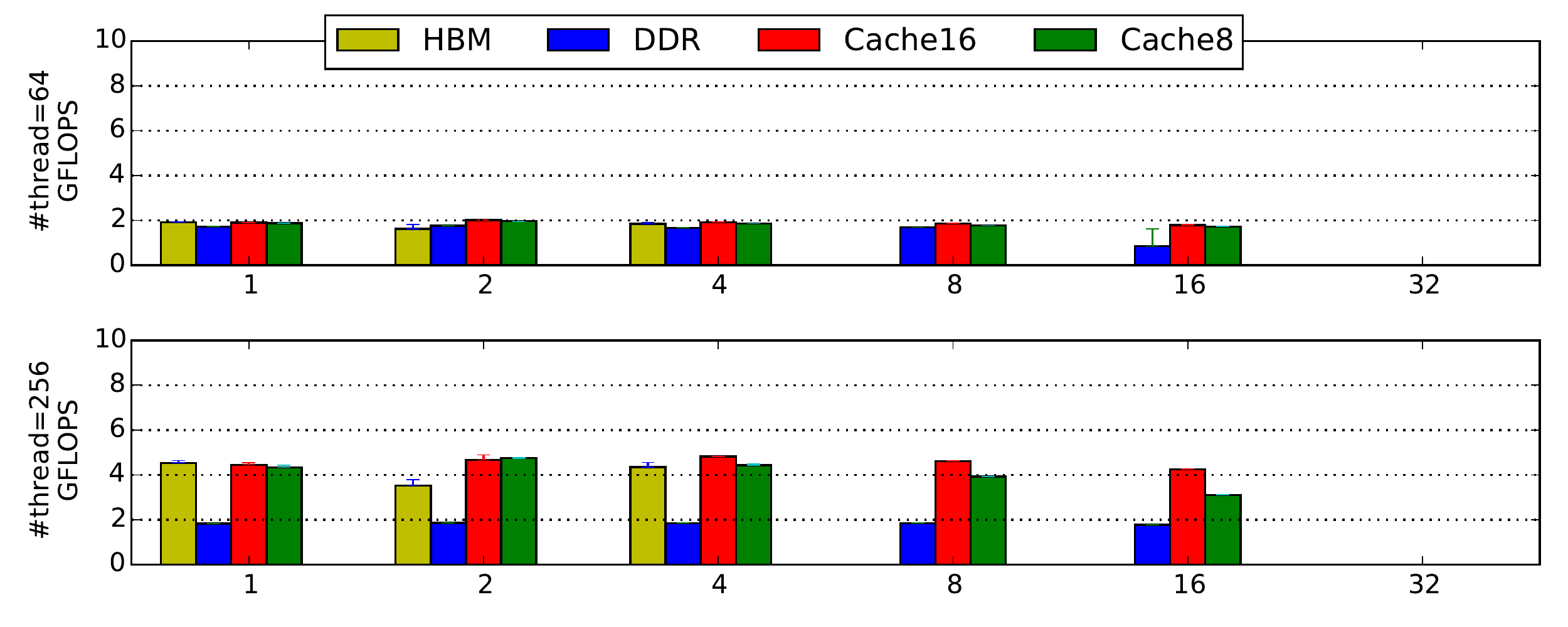}\label{fig:lapra}}
\subfloat[BigStar]{\includegraphics[width=\imscale\columnwidth]{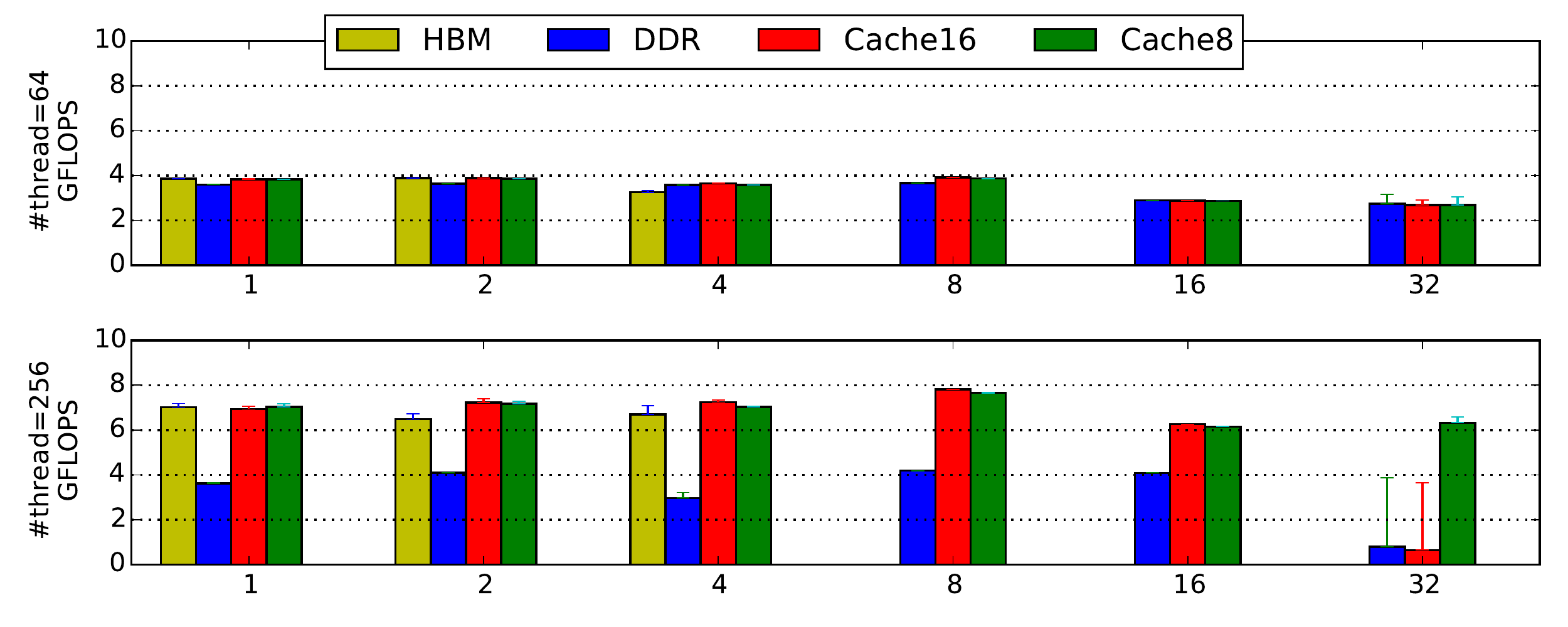}\label{fig:bigra}}

\subfloat[Brick]{\includegraphics[width=\imscale\columnwidth]{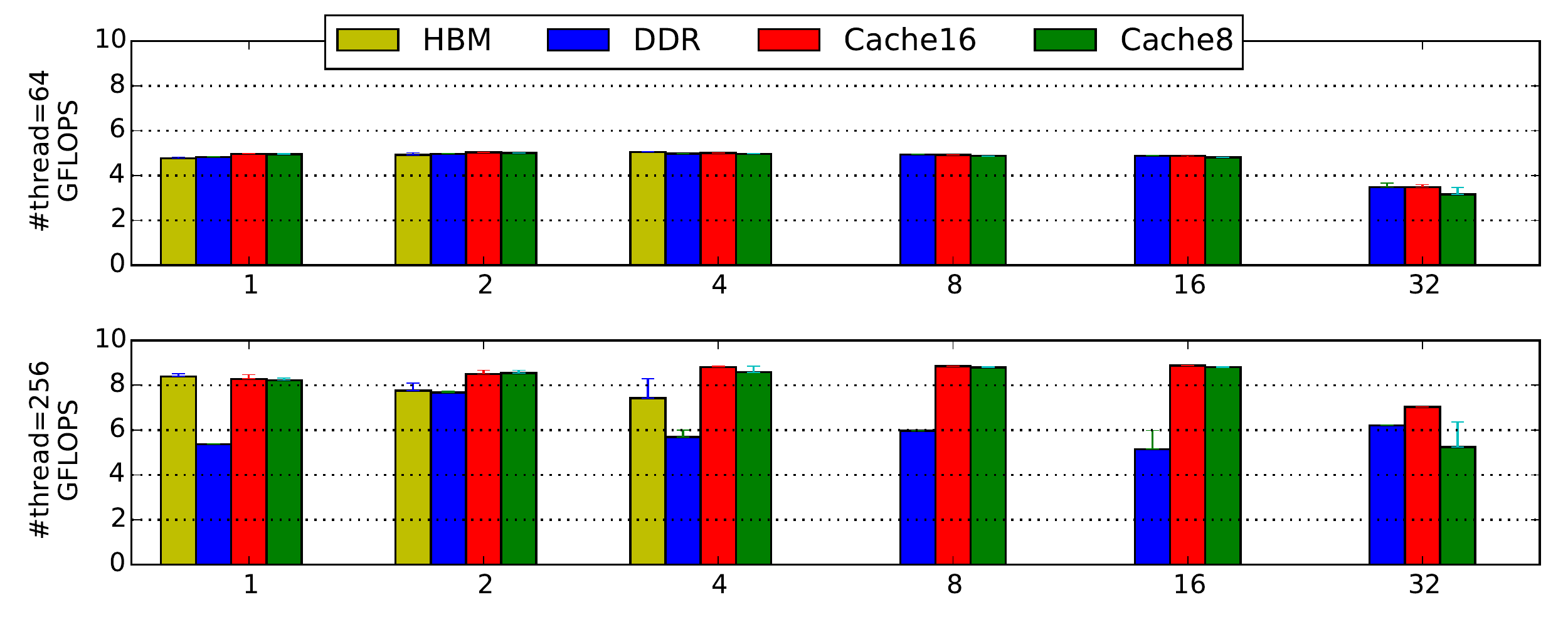}\label{fig:brickra}}
\subfloat[Elasticity]{\includegraphics[width=\imscale\columnwidth]{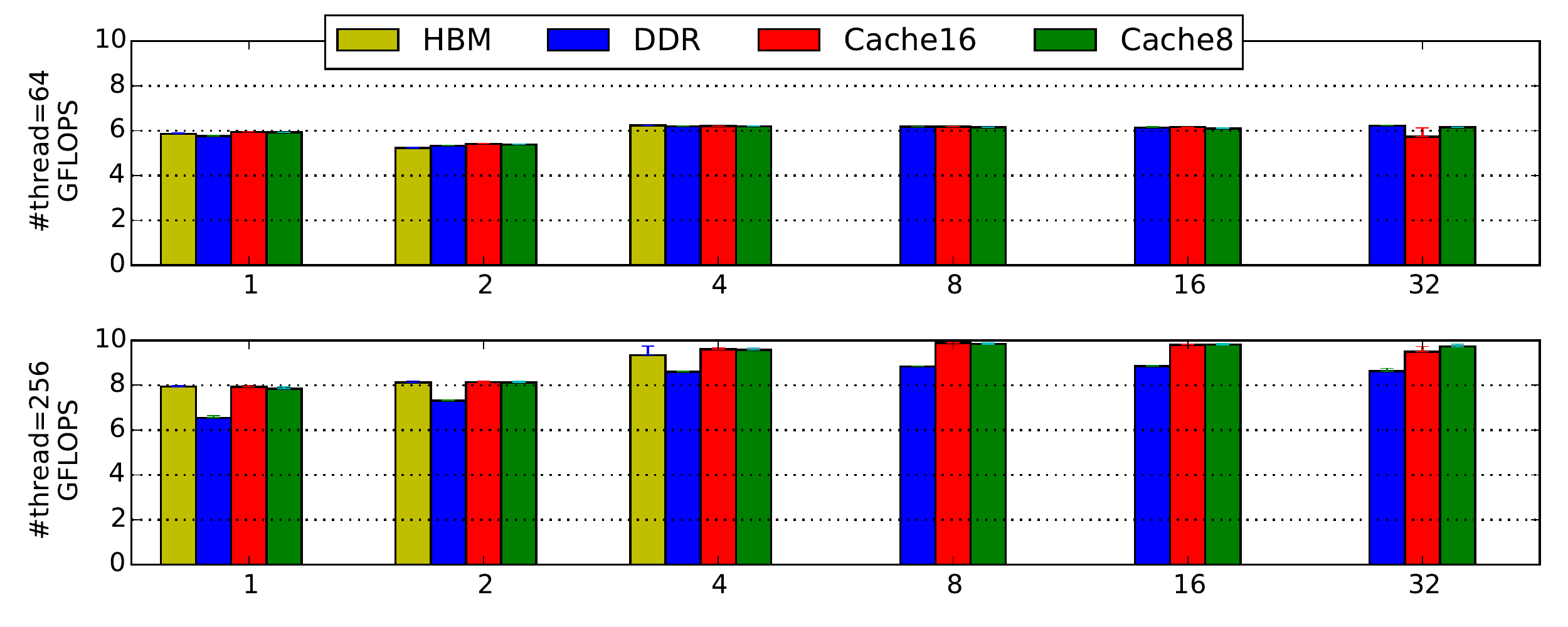}\label{fig:elasra}}
\end{center}
\caption{$R\times A$ GFLOPs on KNL. Top and bottom figures present the GFLOPs achieved on 64 and 256 threads, respectively.
X axis refers to sizes of A matrices (in GB) used in the experiments.
}
\label{fig:knlra}

\end{figure*}

\begin{table}[]
	\centering
	\caption{L2 Cache Miss Percentages for the $R\times A$ and $A \times P$ Problems}
	\label{tab:cache}
	
	\begin{tabular}{l|l|l|l|l}
		& Laplace3D & BigStar & Brick3D & Elasticity \\ \hline
		\multicolumn{1}{l|}{$A \times P$ L2-Miss\%} & $21.52$ & $20.51$ & $8.51$ & $8.23$ \\ \hline
		\multicolumn{1}{l|}{$R \times A$ L2-Miss\%} & $55.07$ & $30.22$ & $13.73$ & $3.20$ \\ 
	\end{tabular}
\end{table}

Using these four instances, we evaluate the
performance of our {\sc kkmem} implementation using the various memory modes supported on
our production Intel Xeon Phi 7250 (\knl{}) systems. These
nodes comprise a single socket 7250 Xeon Phi running at 1.4GHz, which
have 34 dual-core `tiles' (for 68 cores in total), 16GB of on-package
MC-DRAM and 96GB of capacity \ddr{}4 system memory. Each core provides
4-way multi-threading (SMT). We utilize the Intel 17.2
compiler, configured with GCC 4.9.3 for header file and basic system
libraries. The results are presented in Figures~\ref{fig:knlap}
and~\ref{fig:knlra} are for $A\times P$ and $ R\times A$ problems. 

We run these experiments using 64 threads (one thread/core), and 256
threads (with four hyperthreads/core). For each problem domain, we scale
the size of the $A$ matrix from 1GB to 32GB given on the X-axis (the
Laplace problem with 32GB $A$ matrix does not fit into 96 GB \ddr{} and is
excluded). The sizes of $R$ and $P$ are usually much lower and differ
based on the problem (an example for the sizes of $R$ and $P$ can be
found in Table~\ref{tab:gpudp}, the rationale holds for all sizes). We
run both $R\times A$ and $A\times P$ to study the effects of the
structure of the matrices on the left and right hand sides of the
multiplication, excluding $RA\times P$ and $R\times AP$ as the size of the
intermediate results differ for different problems. As a result it is
difficult to run weak scaling tests on these configurations. Moreover,
they usually show the same trends as $R\times A$. Each run is repeated
20 times with the bars representing the median performance. As the
matrix sizes are increased, the overall required size for three matrices
increases to as much as 90\% of the overall memory capacity. At such
scales, the runtimes can fluctuate significantly. We added error bars to
represent the best performance achieved in 20 repetitions, which we
expect to observe in the presence of larger memory capacities.

{\bf \hbm{}} is run using the KNL flat memory mode -- all data is placed
into MC-DRAM (\hbm{}). As the overall \hbm{} capacity is 16 GB, this option is
only run when the overall problem size requires less than 16GB. {\bf
\ddr{}} is also on the same flat memory mode as \hbm{} but in this
configuration all data is placed in \ddr{} memory. {Cache16
(Cache8)} is run on the KNL nodes with the entire (half)  
\hbm{} memory used by the hardware to provide caching of accesses
to the \ddr{} and the rest available as a flat memory space for data
allocation. Even in Cache8 runs we allocate all data
structures in the \ddr{} but use the runtimes to show the effect of a
smaller (8GB) hardware cache. This simulates what would happen when
these kernels are run with real applications.



Figure~\ref{fig:knlap} shows the results for the $A\times P$
multiplication runs. 
We observe that {\sc kkmem} scales beyond the cores with the
hyperthreads for almost all multiplication instances (except the Laplace
$R\times A$ problem in \ddr{}). Its performance is constant across memory
spaces, and it does not take a significant performance change with \hbm{}.
These results suggest that {\sc kkmem} is not bandwidth bounded on \ddr{}
when using 64 threads. We observe benefits of \hbm{} when hyperthreads are
enabled based on the matrix properties. Yet, the performance gain on
$A\times P$ multiplications are rather smaller compared to $R\times A$.
Regular structures of $A$ matrices result in more regular accesses to
$P$, and improve the temporal cache locality. This reduces the global
memory accesses and decreases the algorithms' ability to become
bandwidth bound. As a result, the effect of the use of \hbm{} on $A\times 
P$ multiplication is still minimal. On the other hand, for $R\times A$,
$R$ has more strided accesses and less reuse of the rows of $A$. This
lowers temporal locality, and increases the memory accesses, making the
algorithm more prone to be affected by the bandwidth-related overheads.
Even with $R\times A$, we observe that the performance differences
reduce as we go from Laplace to Elasticity with the increasing $\delta$.
As density of $A$ increases, it benefits from spatial locality as well
as prefetching opportunities. In general, {\em having high temporal or spatial
locality reduce the observed performance difference between \hbm{} and \ddr{}.}

To support this theory, first, we run the
Kokkos-profiling~\cite{edwards2014kokkos} cache measurement tools to measure the L2 cache misses.
These cache-miss ratios are listed in Table~\ref{tab:cache}. In general,
we observe that $A\times P$ multiplications have lower L2 cache-miss rations
than $R\times A$ due to $R\times A$'s poor temporal localities\footnote{
Spatial locality is likely to increase L1 hits, while more temporal
locality is likely to improve L2 hits.}. Larger L2 cache miss ratios cause more 
frequent memory accesses. This exhausts more memory bandwidth and 
results in larger performance differences between \hbm{} and \ddr{}.
One exception to this is in Elasticity, where the performance of $R\times A$ 
is at least as good as $A\times P$ because of the high density of $A$. 

Second, in order to further experiment the effect of the density of 
right handside matrix, we take the $R$ and $A$
matrices belonging to Elasticity problem, and generate right hand-side
matrices for both $R$ and $A$ with increasing $\delta$. For each
problem, we measure the achieved GFLOP/s on the \ddr{} and \hbm{} memories
modes, as well as the L1 and L2 miss ratios using the Kokkos tools cache
profiler. The results are listed in Table~\ref{tab:expdensity}. Similar
to what we have observed, the performance difference between \ddr{} and \hbm{}
for $R\times A$ multiplication decreases with the increasing $\delta$.
As $\delta$ increases, the algorithm makes use of spatial locality more,
which is reflected in the decreasing L1 cache misses. This suggests
that, even when the temporal locality is low, additional spatial
locality reduces global memory accesses, helping the algorithm to be
more resistant to bandwidth changes. We observe smaller performance
differences for $A\times P$. For low $\delta$, there are slight
performance differences, as we observe for our $A\times P$ multiplications 
($\delta$ of $P$ is usually between 3 and 4.5). Note that increasing
$\delta$ is likely to reduce temporal locality as we have seen in
$R\times A$ multiplication, as bigger rows are more likely to increase the
probability of cache eviction.

\begin{table}[t]
\centering
\caption{GFLOP/s achieved on \ddr{} and \hbm{} of KNLs for the multiplications of Elasticity's $R$ and 
$A$ matrices with randomly generated right hand side (RHS) matrices.
L1 and L2 Cache miss ratios are listed on the last two column.}
\label{tab:expdensity}
\resizebox{\tabscale\columnwidth}{!}{
\begin{tabular}{c|r|r|r|r|r}

 & \multicolumn{1}{c|}{$\delta$ of RHS} & \multicolumn{1}{c|}{\ddr{} GFLOP/s} & \multicolumn{1}{c|}{\hbm{} GFLOP/s} & \multicolumn{1}{c|}{L1 M\%} & \multicolumn{1}{c}{L2 M\%} \\ \hline
\multirow{5}{*}{$R \times RHS$} & $1$ & $0.34$ & $0.92$ & $1.1$ & $17.14$ \\ \cline{2-6} 
 & $4$ & $1.09$ & $2.26$ & $0.69$ & $16.45$ \\ \cline{2-6} 
 & $16$ & $2.66$ & $3.59$ & $0.36$ & $11.71$ \\ \cline{2-6} 
 & $64$ & $4.42$ & $4.78$ & $0.27$ & $8.06$ \\ \cline{2-6} 
 & $256$ & $4.79$ & $4.99$ & $0.29$ & $6.31$ \\ \hline
\multirow{5}{*}{$A \times RHS$} & $1$ & $1.17$ & $1.69$ & $1.09$ & $3.96$ \\ \cline{2-6} 
 & $4$ & $3.00$ & $3.27$ & $0.89$ & $3.14$ \\ \cline{2-6} 
 & $16$ & $4.19$ & $4.32$ & $0.46$ & $3.98$ \\ \cline{2-6} 
 & $64$ & $5.00$ & $5.02$ & $0.3$ & $5.1$ \\ \cline{2-6} 
 & $256$ & $5.07$ & $5.14$ & $0.29$ & $6.1$ \\ \hline
\end{tabular}
}
\end{table}

Another observation from Figure~\ref{fig:knlap} and~\ref{fig:knlra} is
that the KNL caching-modes achieve as good performance as with \hbm{}.
While on first thought this is trivial to achieve when the whole
problem fits into \hbm{}, it reflects the efficient design of the hardware
even when additional caching logic is active. In this experiment, we
also study the performance of the cache-mode by increasing the problem
sizes, and reducing the effective cache sizes on the KNL
processor. We observe that all cache-modes are mostly successful at maintaining the
performance achieved by \hbm{} on larger scales.
We observe a slight performance
difference on Laplace $R\times A$ multiplication, the problem with the
worst spatial and temporal locality, between lower cache sizes. However,
we conclude that even a small cache-size as low as 8GB is sufficient to
achieve \hbm{} performance on large scale problems.


On memory systems differing mainly for bandwidths such as MC-DRAM and
\ddr{}, we do not observe high performance differences for {\sc kkmem}.
{\em The performance difference of \ddr{} and \hbm{} decreases with 
better temporal and spatial locality}. Moreover, the vendor-provided 
cache mode achieves the performance of \hbm{} even for the cases where 
the problem size is much larger than the cache size. {\em This suggests
a reduced need of a multi-level SpGEMM algorithm for
memory subsystems where the main differences are in terms of bandwidth.}

\subsubsection{Selective Data Placement}

As shown in the previous section cache-mode maintains the performance of
\hbm{} on large scale problems. However, cache-mode is a boot-time (BIOS)
option for the compute node. Large-scale KNL systems are known to
experience significant node reconfiguration times. Additionally,
applications which utilize sparse-matrix multiplication will often use
large portions of the system memory for other application data or
problem state. In such cases, the application may run in flat mode with
a problem size that does not fit into \hbm{}. Anecdotal experience from our
production environment at Sandia has shown this mode of execution to be
common enough for developers to request support in the Trilinos
framework. In these cases, {\sc kkmem} performance in \ddr{} can be as low
as half of the performance of \hbm{} (Laplace $R\times A$). We can perform a selective data
placement (deciding where to allocate specific data structures), where
only one or two matrices are stored in \hbm{}. As explained in previous
sections, for $C=A\times B$, the accesses to $A$ and $C$ are regular and occur in
a streaming fashion in {\sc kkmem}. Moreover, since we utilize
sparse accumulators, the accesses to these hashmap accumulators are
mostly localized in caches. $A$, $C$, and the accumulators are not
likely to need higher bandwidth. On the other hand, the accesses to $B$
can be irregular depending on the structure of $A$. As a result, for the
cases where \hbm{} cannot store the entire problem, we propose storing only
B in \hbm{} to recover \hbm{} performance. We show the effect of this
method in the experiments section. This method, DP (data placement),
only works when $B$ fits into \hbm{}. 

\subsubsection{Chunking Method For KNLs}

\begin{figure}
	\begin{center}
		\includegraphics[width=\imscale\columnwidth]{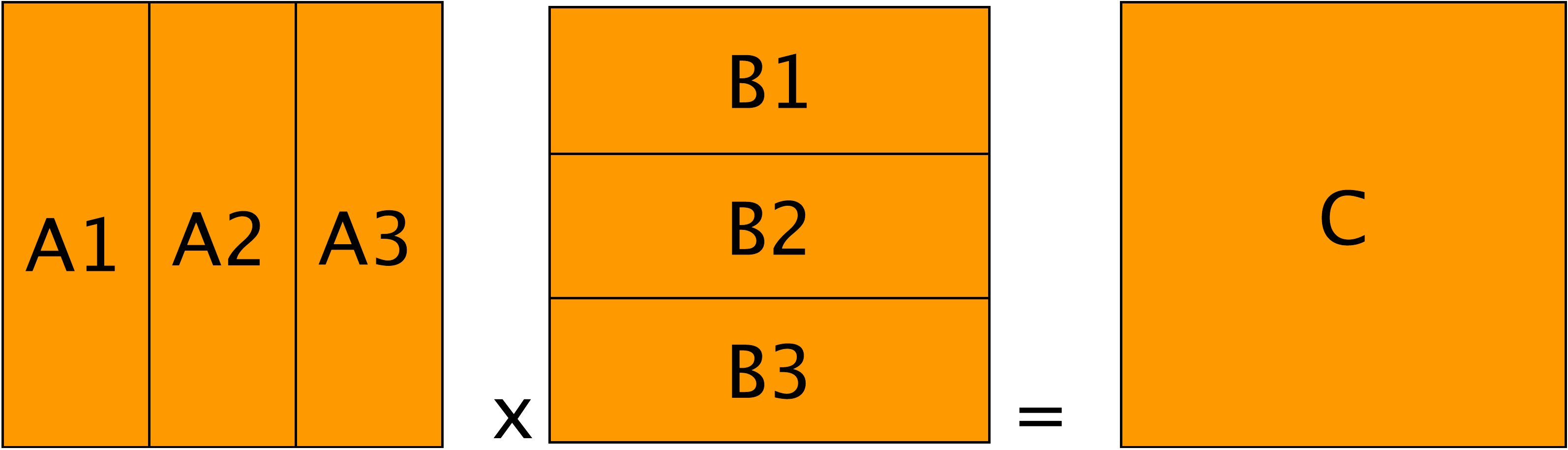}
	\end{center}
	\caption{Example of a simple chunking for KNL}
	\label{fig:knlpart}
\end{figure}

Problems where $B$ is larger than \hbm{} requires partitioning of $B$. 
Column-wise partitions 
have been explored in one level memory before ~\cite{patwary2015parallel}.
However, since our data is stored row-wise, finding column-wise
partitions that will fit into \hbm{} is usually prohibitively expensive.
Instead, we use a row-wise partition of $B$ as shown in the
Figure~\ref{fig:knlpart}. Assume $B$ is partitioned into three chunks, with
each chunk fitting into \hbm{}. Note that, a row-wise partition of $B$
induces a column-wise partition of $A$ as in the figure, as rows of
$B_1$ are only accessed by the columns in $A_1$. We avoid explicit
column-wise partition of $A$ because of the introduced partitioning
overhead in practice. Instead, the multiplication kernel is provided with the
row ranges of the $B$ partition, allowing it to skip any columns of $A$
outside of this range (we do not assume that columns are sorted). A
slightly modified {\sc kkmem} kernel is used as a subprocedure to
perform a matrix multiplication with row ranges. This kernel first performs $A_1 \times
B_1$ to find the partial result $C^1$. Then $C^2$ is found by performing
a multiplication and a matrix addition as $C^2 = A_2 \times B_2 + C^1$.
This subprocedure is a fused multiply and add kernel. Once a
multiplication for row is completed, it inserts the existing values of
$C^1$ into its hashmap accumulators to find $C^2$. 
Algorithm~\ref{alg:chunkknl} shows the simple chunking strategy for KNLs.
First, the number of partitions are found so parts of $B$ fits in memory. With a binary search, 
we find $P_B$ which stores the begin and end row indices of
each partition. Then one by one, each row partition of $B$ 
(ranges are defined in $B_{rp}$) is copied into fast memory and 
a fused multiple/add {\sc kkmem} is used to compute $C$.
As observed, there is little room for improvement between \ddr{} and \hbm{} on KNLs
in most cases. Hence, we expect the amortization of data movement cost of chunking in 
only a few cases.

\begin{algorithm}[htb]
\smallerfont
\begin{algorithmic}[1]
\Procedure{kkmemKNLchunk}{$A,B,C,FastSize$}

  \State $np = ceil(\frac {size(B)} {FastSize})$
  \State $pSize = \frac{size(B)} {np} $
  \State $P_{B} = BinarySearch(B,pSize)$

  \ForAll{$B_{rp} \in P_{B}$}
    \State $FastB =copy2Fast(B, B_{rp})$ 
    \State $C = kkmem(A, FastB, C, B_{rp})$
  \EndFor
\EndProcedure
\end{algorithmic}
\caption{Chunking method for KNL. FastSize refers to \hbm{} size.}
\label{alg:chunkknl}
\end{algorithm}

\subsection{Analysis on GPUs}

We perform the experiments on the same matrices using NVIDIA's P100
(``Pascal'') GPUs. Each GPUs has a dedicated high bandwidth global
memory, which we refer as \hbm{}. In our system, each P100 is connected to
the host IBM OpenPOWER8 host-processor architecture with NVLINK (Version
1)~\cite{foley2017ultra}. The GPU can access the host CPU pinned memory
space directly. Accessing pinned memory has high latency overheads, as
well as much lower bandwidth. This is different from \knl{}s
where \hbm{} and \ddr{} have approximately similar latency overheads.
Moreover, NVLINK additionally provides unified memory spaces (\uvm{}) that
handle data movements automatically between host and GPU memories. A
data structure that is allocated through \uvm{} can be located in \hbm{} or
pinned memory based on the request from host and GPU sides. As a result,
\uvm{} works in a similar way to cache-mode in \knl{}s.  




\begin{figure*}
\begin{center}
\subfloat[Laplace]{\includegraphics[width=\imtscale\columnwidth]{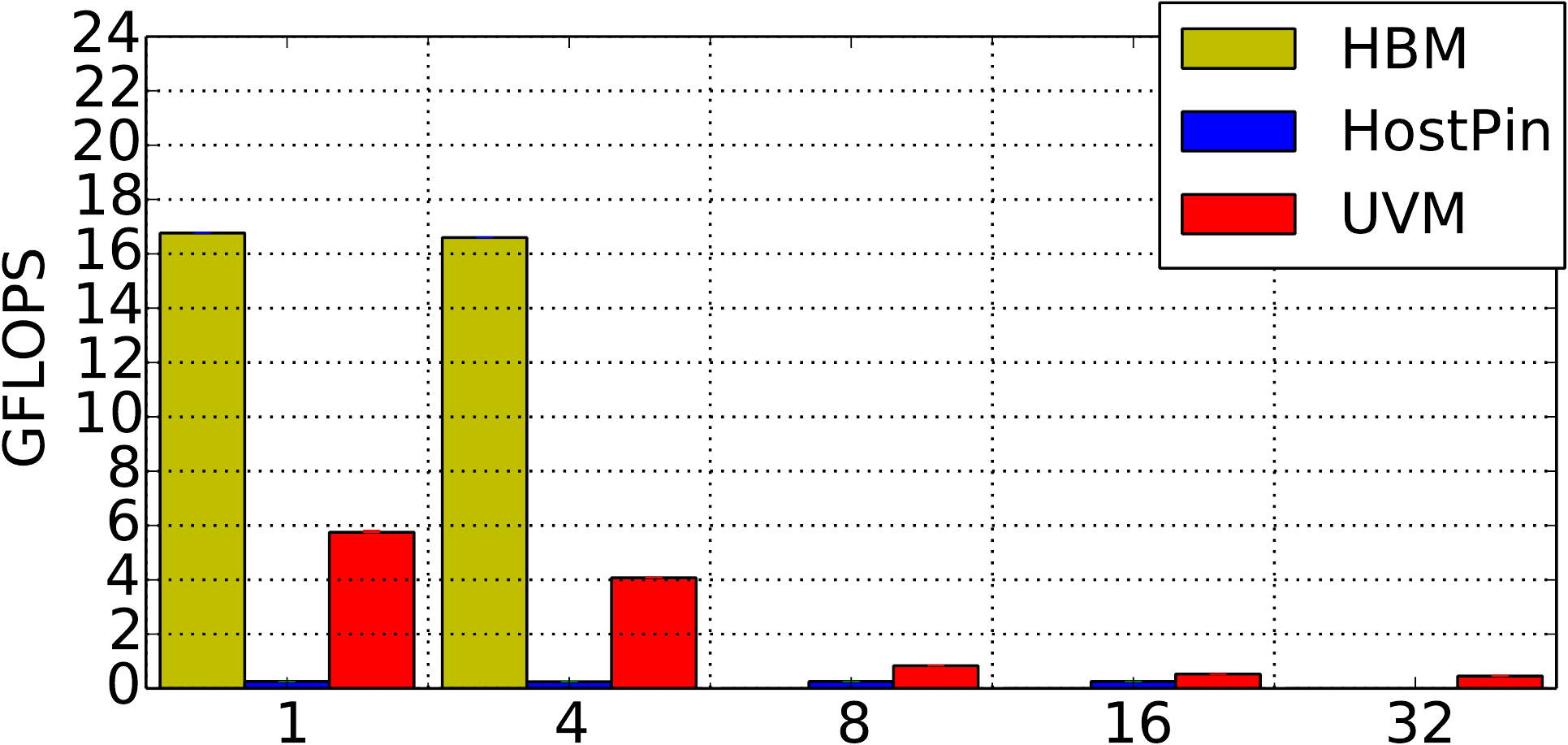}\label{fig:gpulapap}}
\subfloat[BigStar]{\includegraphics[width=\imtscale\columnwidth]{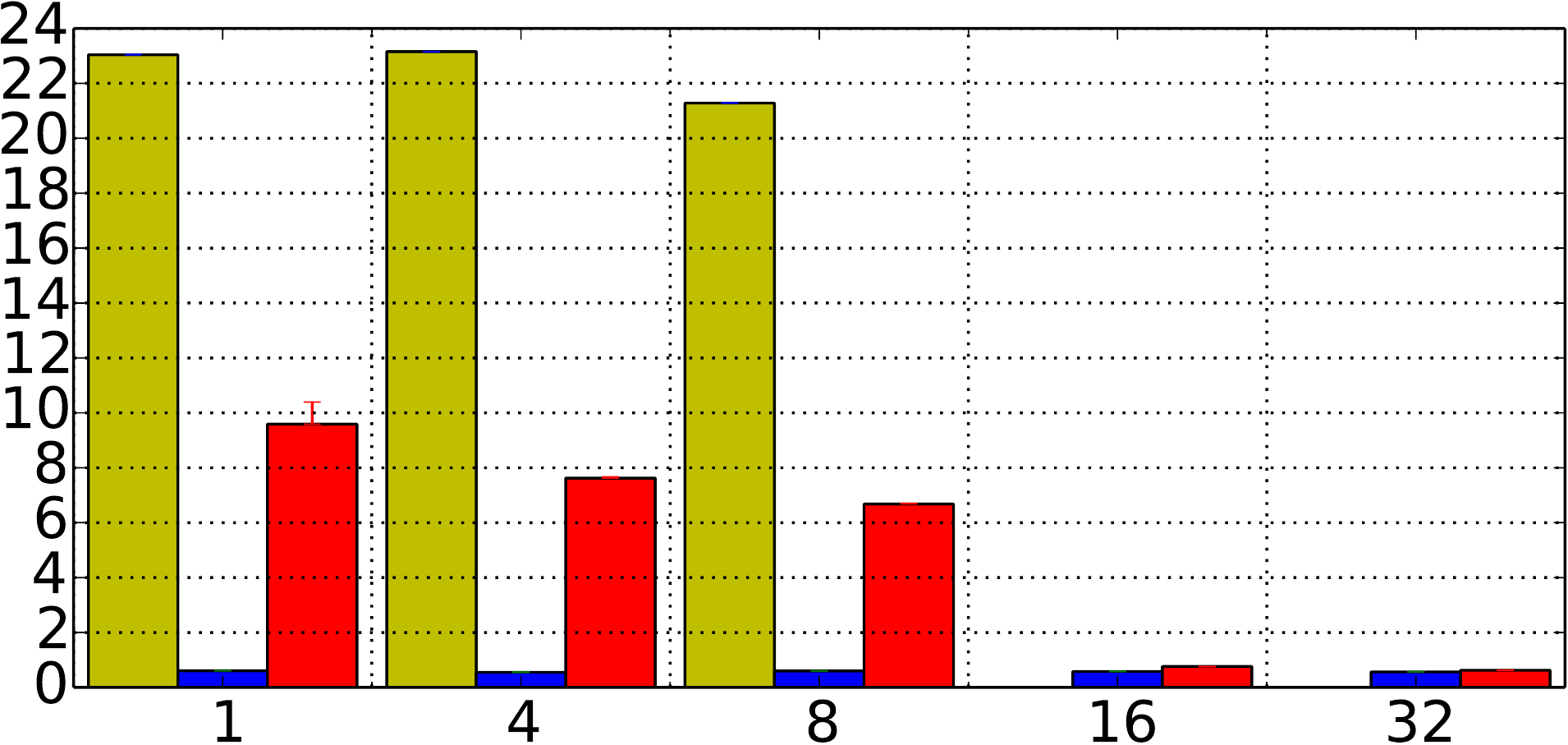}\label{fig:gpubigstarap}}

\subfloat[Brick]{\includegraphics[width=\imtscale\columnwidth]{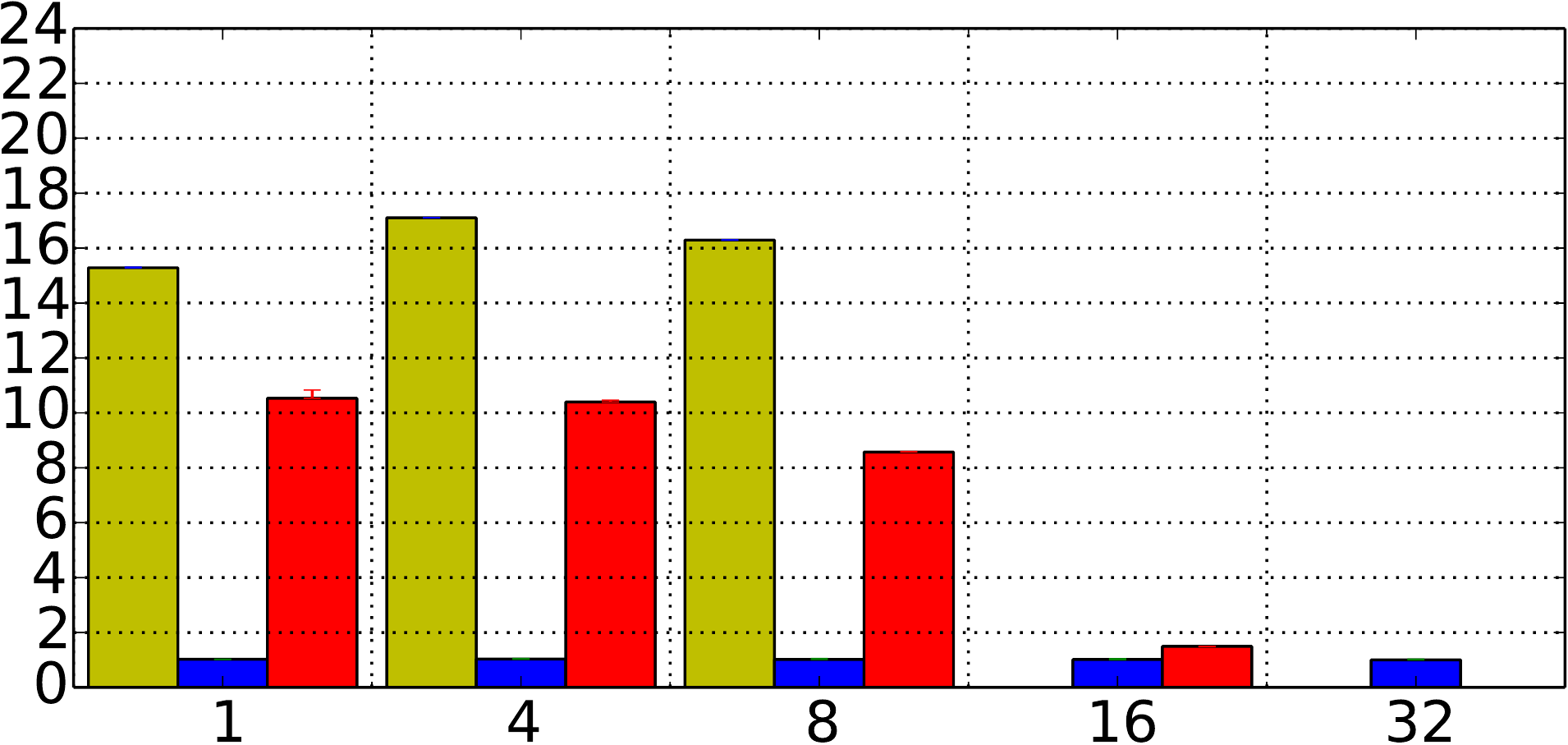}\label{fig:gpubrickap}}
\subfloat[Elasticity]{\includegraphics[width=\imtscale\columnwidth]{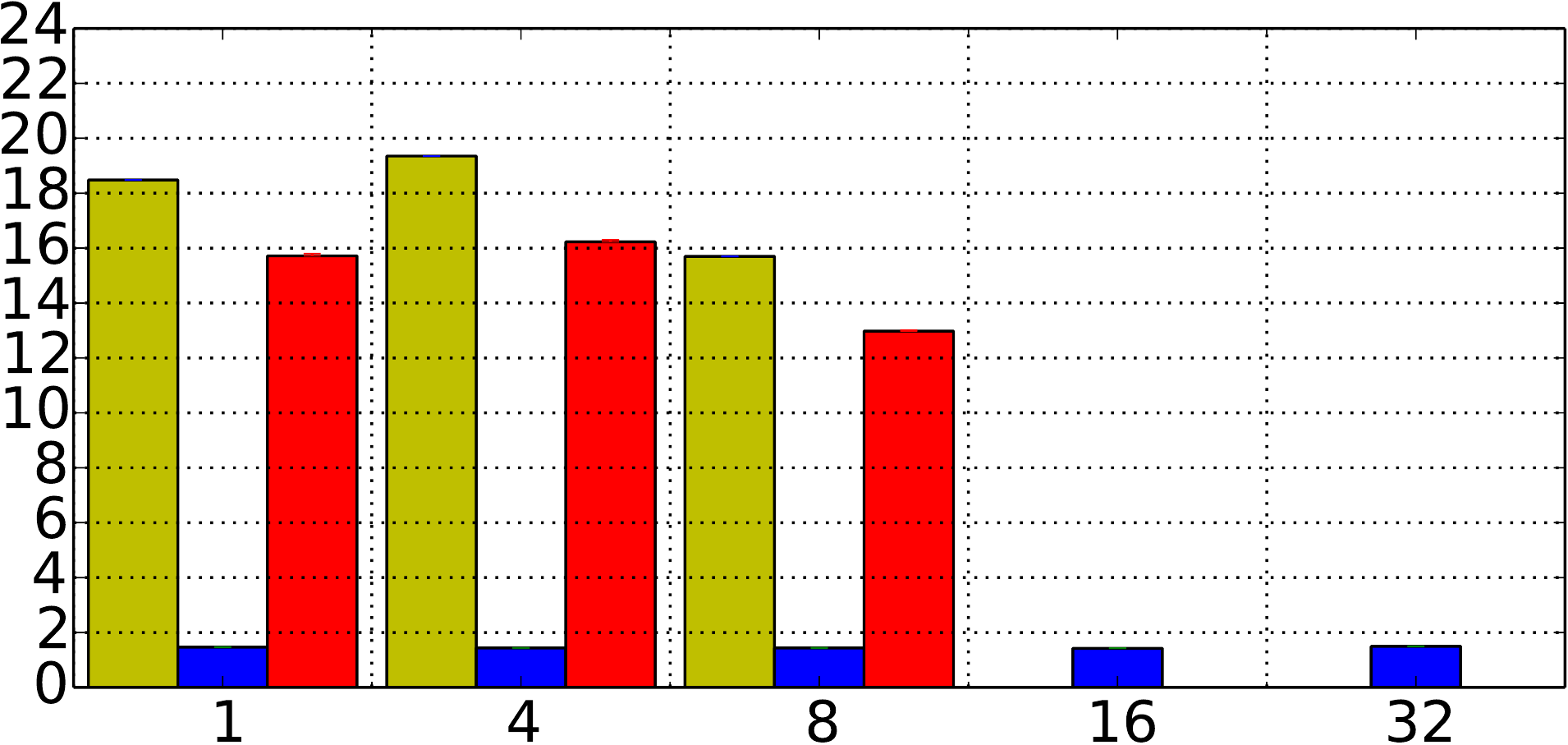}\label{fig:gpuelasticityap}}
\end{center}
\caption{AxP
$A \times P$ GFLOPs achieved by \hbm{}, Pinned Memory, and \uvm{} on P100 GPUs. X axis refers to sizes of A matrices (in GB) used in the experiments.
}
\label{fig:gpuap}
\end{figure*}

\begin{figure*}
\begin{center}
\subfloat[Laplace]{\includegraphics[width=\imtscale\columnwidth]{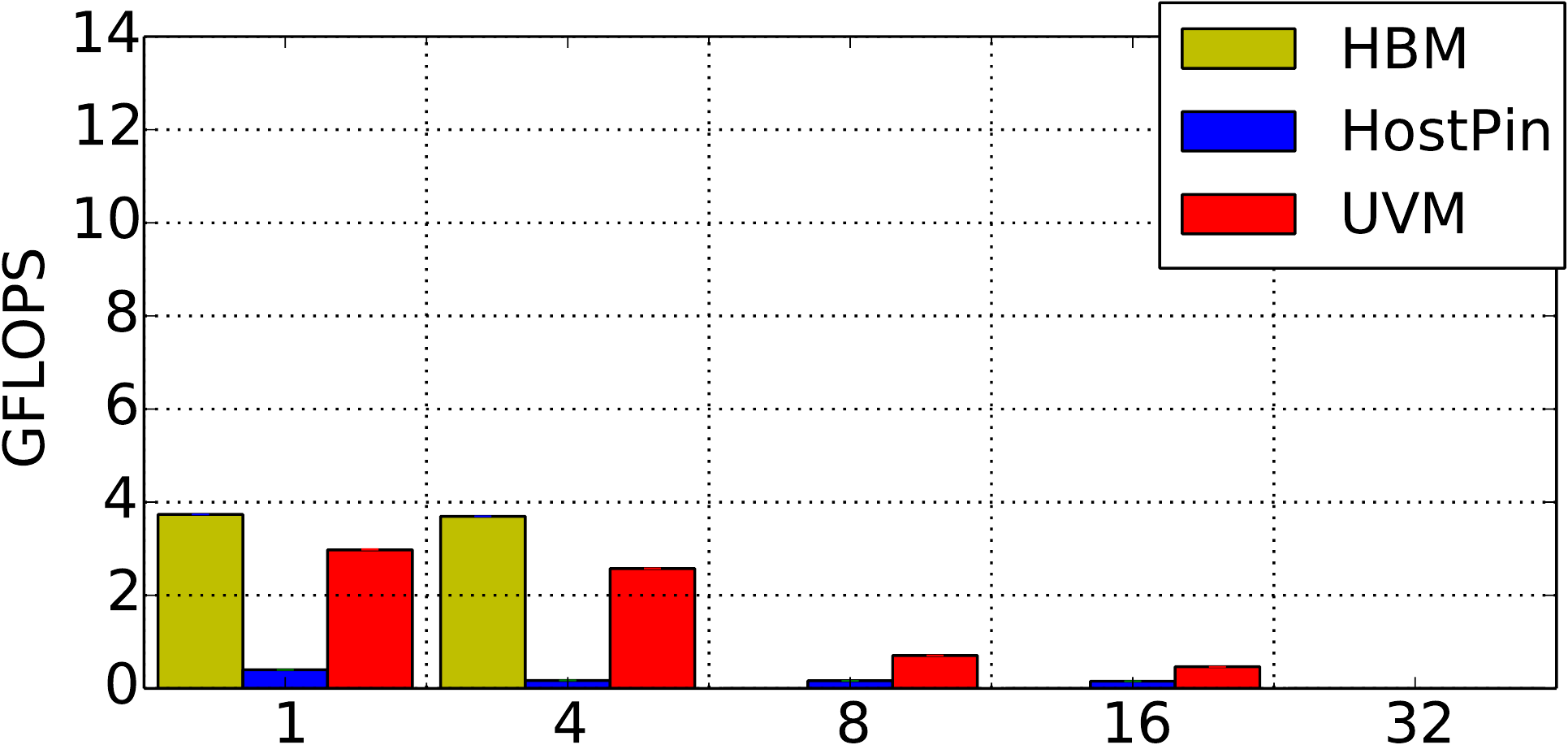}\label{fig:gpulapra}}
\subfloat[BigStar]{\includegraphics[width=\imtscale\columnwidth]{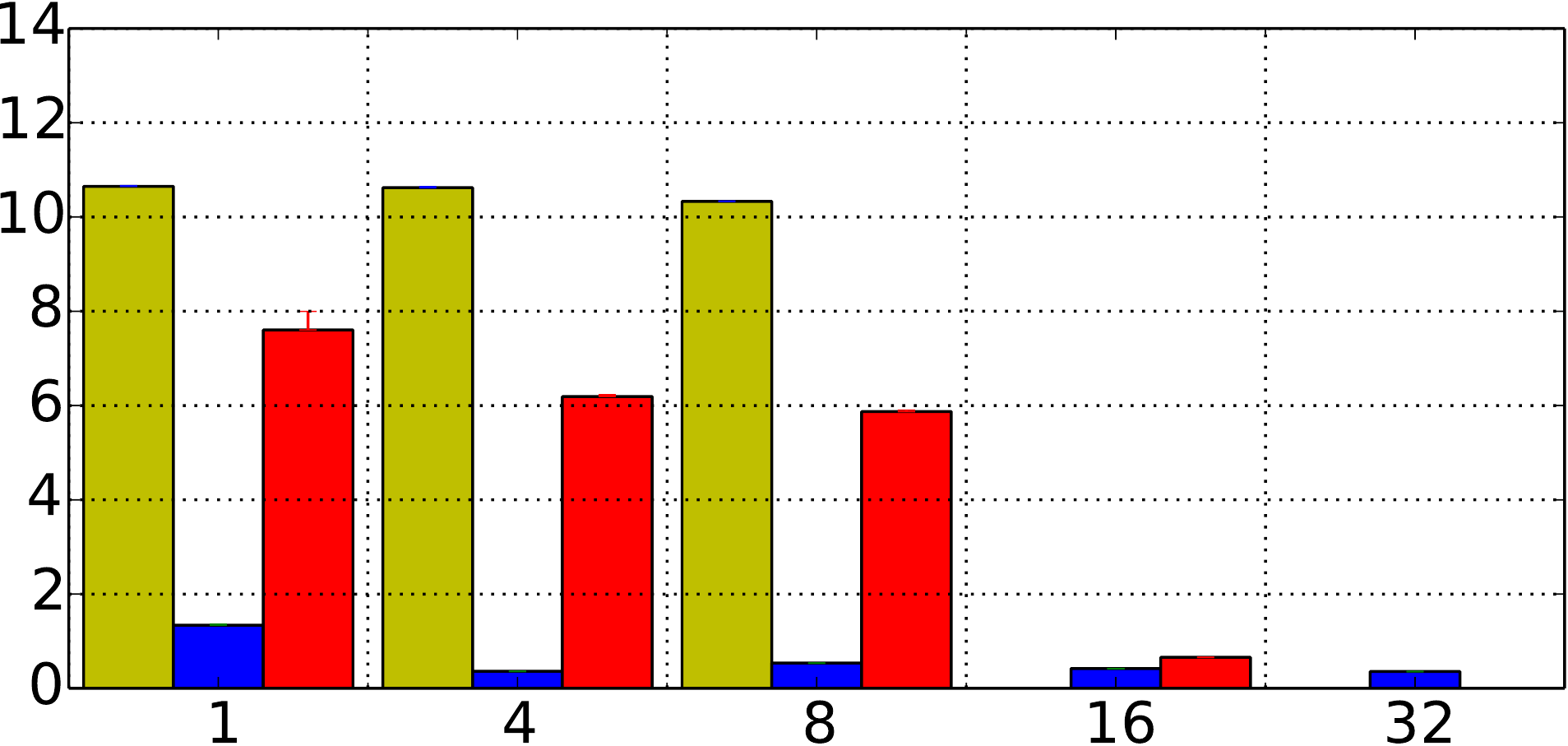}\label{fig:gpubigstarra}}

\subfloat[Brick]{\includegraphics[width=\imtscale\columnwidth]{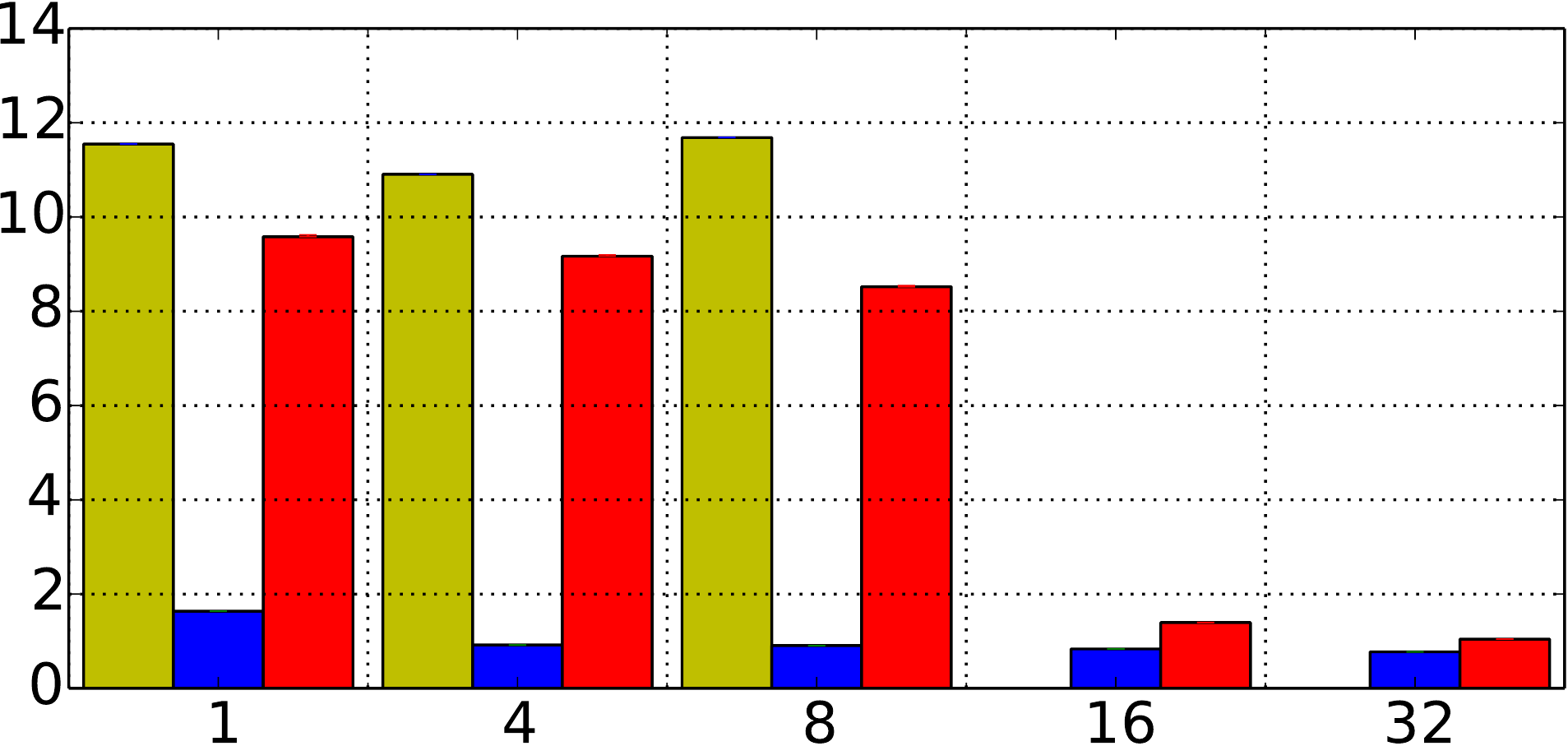}\label{fig:gpubrickra}}
\subfloat[Elasticity]{\includegraphics[width=\imtscale\columnwidth]{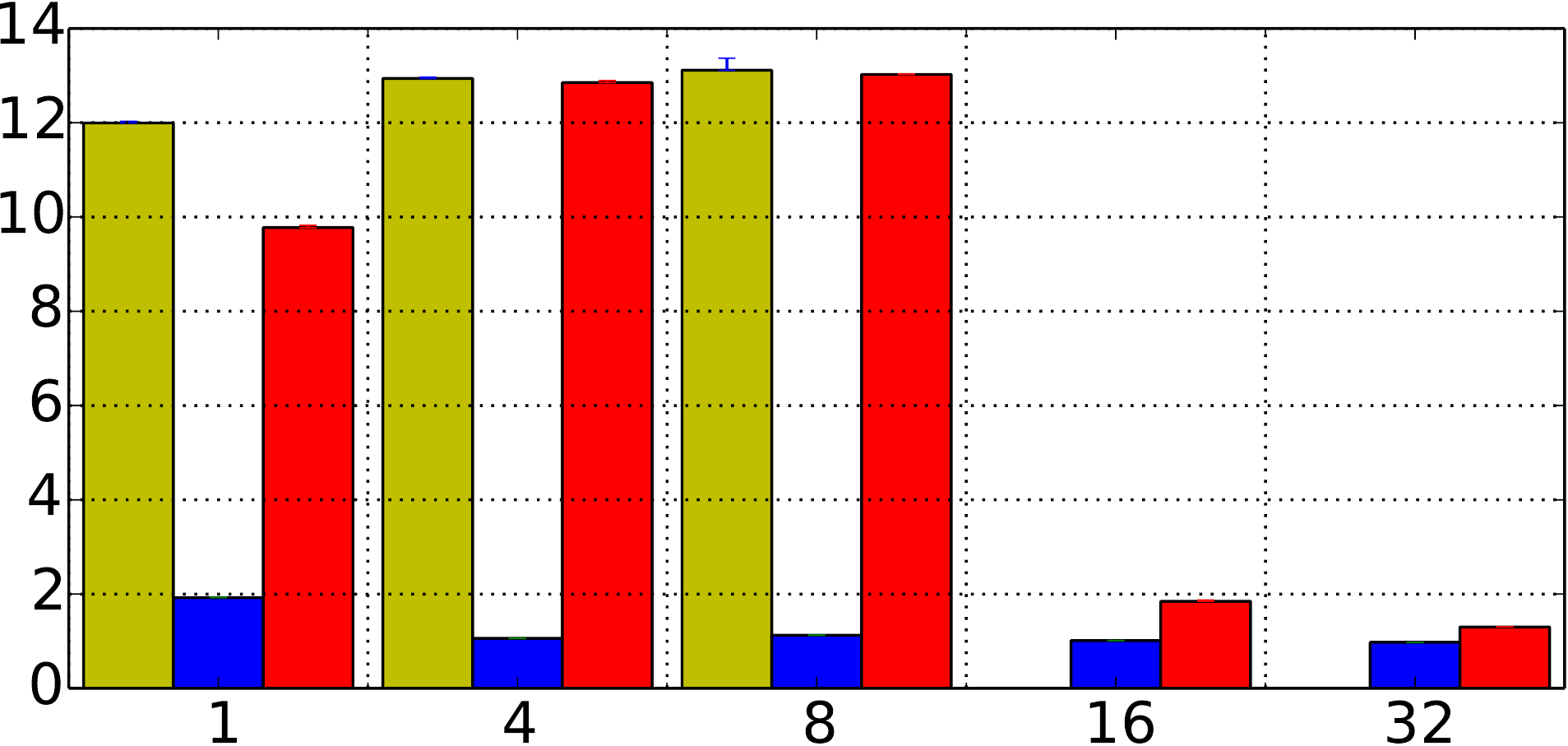}\label{fig:gpuelasticityra}}
\end{center}
\caption{
$R \times  A$  GFLOPs achieved by \hbm{}, Pinned Memory, and \uvm{} on P100 GPUs. X axis refers to sizes of A matrices (in GB) used in the experiments.
}
\label{fig:gpura}
\end{figure*}

Using these three configurations, we perform the same multiplications as earlier.
Results in Figure~\ref{fig:gpuap} and~\ref{fig:gpura} show
achieved GFLOP/s are usually much higher on the \gpu{} systems than \knl{}s. We observe much
higher performances for $A\times P$ w.r.t. to $R\times A$. Both spatial
and temporal locality on \knl{}s translate to coalesced memory accesses on
\gpu{}s. Based on the structure of the matrices, a half, quarter, {\em etc.},
warp may be assigned for the computation of a single row. When a warp is
partitioned across the consecutive rows, the structure of $A$ improves
coalesced memory accesses since consecutive rows of $A$ are likely to
have closer or even the same memory accesses to $P$. Moreover, our algorithm
allocates its first level hashmap accumulators in the \gpu{} shared memory.
When the values do not fit into first level hashmap, the second level is
allocated in the \gpu{}'s global memory. $A\times P$ matrix multiplications tend
to have fewer nonzeroes in the result. As a result, most of the hashmap
insertions happen in the faster \gpu{} shared memories.

We observe a huge performance drop when host pin
memory is used. This suggests that, although {\sc
kkmem} is tolerant to bandwidth drops, it is much more affected by
significant memory latency overheads. We expect to see similar differences for
wide latency disparity in systems such as non-volatile memory with \ddr{} or \hbm{}. 

For problems that fit into the \gpu{} global memory, \uvm{} should not have
any overheads, as the data  moved to \gpu{}s are never brought back to
host memory. Therefore, \uvm{} should not have data movement cost in and
out of \gpu{}s during execution. We observe that \uvm{} achieves as low as
$30\%$ of the \hbm{} performance for such cases. We observe bigger
performances drops with \uvm{} w.r.t. \hbm{} when the problem size gets
larger. Whenever the problem requires more memory than \hbm{}, \uvm{} is
observed to achieve only the performance of pinned memory. The missing
data points in the chart are because the method did not complete within
the 2000 second limit applied to our benchmark runs, which is the case
for \uvm{} for multiple instances. The penalty that is paid to access a data 
that is not in \hbm{} is much higher on \gpu{}s than \knl{}s. Therefore,
a general caching method, such as \uvm{}, cannot guarantee smart prefetches without
any knowledge about what memory locations that the method is going to require.
On the basis of our benchmarked results, we conclude that \spgemm{} can
benefit from multi-level chunking algorithm on \gpu{} systems that have 
memory spaces with differing latencies. 

\subsubsection{Chunking Algorithm on \gpu{}s}

The huge performance difference between pinned memory and \hbm{}, as well as
the poor performance of \uvm{} on large matrices, makes a chunking algorithm
vital for \gpu{}s. The chunking algorithm that is used for \knl{}s can
still be used; however because of the different characteristics of the
memory systems, a different chunking strategy is required. For the design
decisions, we perform an experiment given in Table~\ref{tab:gpudp}. In
this experiment, we first test the effect of placing different matrices
in either \hbm{} or host pin memory. \hbm{} refers to the performance of {\sc kkmem}
when all data is in \hbm{}, and HostPin refers to its performance when all
data is in host-pinned memory. For the other columns, we place one of
$A$, $B$ and $C$ ($A\times B= C$) into host-pinned memory while the others are kept in \hbm{}.
The sizes (in GBs) of these matrices are given on the right side of the
table. When $B$ is placed into host-pinned memory, the performance drop ranges
from 7$\times$ to 29$\times$. In these examples, either the sizes of $B$
are large, or they are accessed frequently. As a result, leaving $B$ in
host-pinned memory is not a performant option. On the other hand, the effects of placing
$A$ and $C$ differs based on the different problems. The effect of
placing these matrices into host-pinned memory is minimal whenever the sizes as
well as the accesses to these matrices are much lower compared to B. For
example in Elasticity problem with $R\times A=RA$, the sizes of $R$ and $RA$ 
are 5\% and 11\% of the total problem size. Moreover, each row of $A$ is accessed
roughly 4.5 times. In this case the accesses to $R$ and $RA$ are around
1\% and 3\% of the overall memory accesses. On the other hand, when performing
Laplace $R\times A$, these ratios and their effects on performance are much higher. 
The effect of the placement is also high for $A\times P$ multiplication,
as $A$ is the largest matrix, and accesses to $A$, $B$ and $C$  are more 
uniform.

\begin{table}[t]
\centering
\caption{Achieved GFLOPs with the different data placements on GPUs.}
\label{tab:gpudp}
\resizebox{\tabscale\columnwidth}{!}{
\begin{tabular}{cc|r|r|r|r|r||r|r|r}
 &  \multicolumn{1}{c}{} & \multicolumn{5}{c||}{GFLOPS} & \multicolumn{3}{c}{Size (GB)} \\ \cline{3-10} 
 &  & \multicolumn{1}{c|}{HBM} & \multicolumn{1}{c|}{A\_Pin} & \multicolumn{1}{c|}{B\_Pin} & \multicolumn{1}{c|}{C\_Pin} & \multicolumn{1}{c||}{HostPin} & \multicolumn{1}{c|}{A} & \multicolumn{1}{c|}{B} & \multicolumn{1}{c}{C} \\ \cline{3-10}

\multicolumn{1}{c|}{\multirow{2}{*}{Laplace}} & RxA & $3.68$ & $2.98$ & $0.17$ & $1.91$ & $0.15$ & $2.3$ & $4$ & $5$ \\ \cline{2-10} 
\multicolumn{1}{c|}{} & AxP & $16.67$ & $3.68$ & $1.56$ & $0.26$ & $0.21$ & $4$ & $2.3$ & $5$ \\ \hline
\multicolumn{1}{c|}{\multirow{2}{*}{BigStar}} & RxA & $10.65$ & $9.38$ & $0.36$ & $3.09$ & $0.30$ & $1.5$ & $6.6$ & $3$ \\ \cline{2-10} 
\multicolumn{1}{c|}{} & AxP & $23.20$ & $2.26$ & $2.95$ & $0.65$ & $0.54$ & $6.6$ & $1.5$ & $3.3$ \\ \hline
\multicolumn{1}{c|}{\multirow{2}{*}{Brick}} & RxA & $11.11$ & $10.95$ & $0.94$ & $11.34$ & $0.18$ & $0.5$ & $4$ & $1.8$ \\ \cline{2-10} 
\multicolumn{1}{c|}{} & AxP & $17.10$ & $4.15$ & $1.45$ & $2.24$ & $0.37$ & $4$ & $0.5$ & $1.8$ \\ \hline
\multicolumn{1}{c|}{\multirow{2}{*}{Elasticity}} & RxA & $12.94$ & $12.75$ & $1.08$ & $12.89$ & $0.44$ & $0.25$ & $3.9$ & $0.5$ \\ \cline{2-10} 
\multicolumn{1}{c|}{} & AxP & $19.34$ & $3.71$ & $2.19$ & $4.87$ & $0.50$ & $3.9$ & $0.25$ & $0.5$ \\ 
\end{tabular}
}
\end{table}

\begin{figure}[t]
\begin{center}

\includegraphics[width=\imscale\columnwidth]{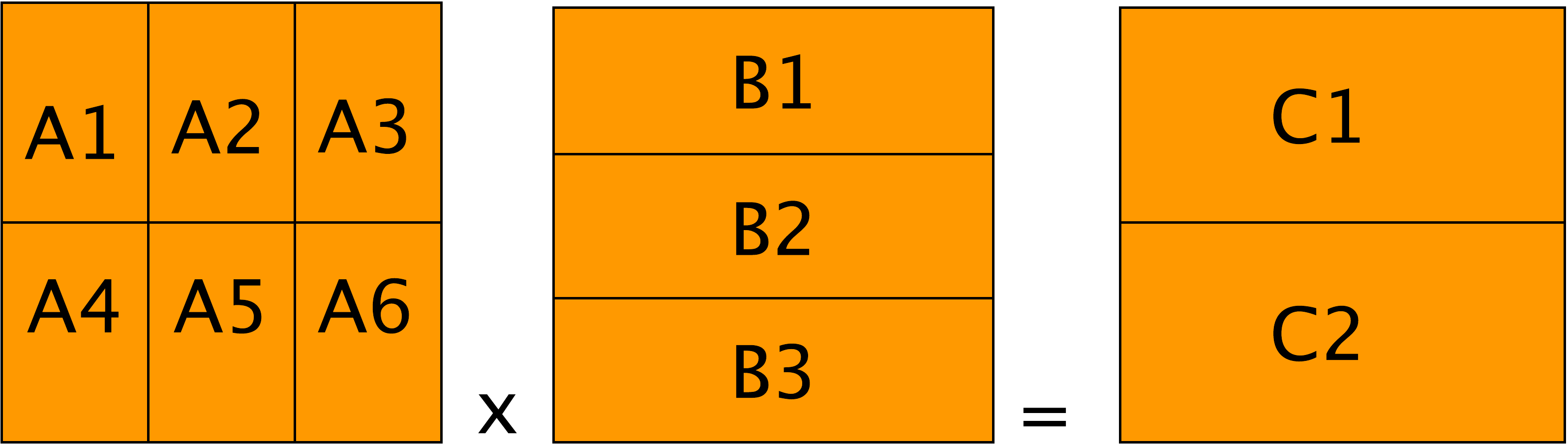}
\end{center}
\caption{Chunking Example for GPUs}
\label{fig:gpupart}
\end{figure}

Based on these results, in our chunking algorithm, we chunk all of the
$A$, $B$ and $C$ matrices. 
Note that, there are cases in which
keeping $A$ or $C$ in pinned memory does not harm the performance. These
are the cases when the sizes of $A$ and $C$ are found to be very small.
As a result the cost of the data movements are also expected to be very
low.
Figure~\ref{fig:gpupart} shows a simple chunking method for \gpu{}s. In
addition to row-wise partitions of $B$ ($P_B$), now we also have row-wise
partitions of $A$ and $C$ ($P_{AC}$). Note that, as in \knl{}s, we do not physically
partition the $A$ column-wise; instead multiplication kernel skips
columns of $A$ based on the row range of $B$. For such 2D partitioning
algorithm, two different streaming orders of chunks can be followed. For example
{\sc kkmemGPUchunk1}, in Algorithm~\ref{alg:chunk1}, brings a row-wise
partition of $A$ and $C$ into fast memory. This partition of $A$ and $C$
is kept in fast memory and partitions of $B$ are streamed to fast memory
and multiplied. The algorithm first performs the partial multiplication 
$C^1_1 = A_1 \times B_1$ ($C^1_1$ denotes partial result for $C_1$), 
and then brings the next chunk of $B$, $B_2$. It then performs 
$C^1_2 = A_2 \times B_2 + C^1_1$.


\noindent
The outer loop in the algorithm copies a row partition of A and C into
fast memory. Note that $C$ is initially empty; only its row pointers
need to be copied from the slow memory. The inner loop brings different row
partitions of B into fast memory. Each iteration of the inner loop
calculates a partial result for corresponding rows of $C$. As a result,
when the inner loop terminates, a final product is found for the
partition of $C$ and this is copied back to slow memory. Then outer loop
brings the next $A$ and $C$ into fast memory. The algorithm copies $A$
and $C$ once, while $B$ chunks are copied as many as the number of
row partitions of $A$ and $C$. The copy cost of this algorithm becomes
$size(A) + size(C) + size(B) \times \|P_{AC}\|$.

\begin{algorithm}[t]
\smallerfont
\begin{algorithmic}[1]
\Procedure{{\sc kkmemGPUchunk1}}{$A,B,C, P_{AC}, P_{B}$}

\ForAll{$AC_{rp} \in P_{AC}$}
  \State $FC =copy2Fast(C, AC_{rp})$
  \State $FA =copy2Fast(A, AC_{rp})$

  \ForAll{$B_{rp} \in P_{B}$}
    \State $FB =copy2Fast(B, B_{rp})$ 
    \State $FC = kkmem(FA, FB, FC,AC_{rp}, B_{rp})$
  \EndFor
  \State $copy2Slow(FC, C, AC_{rp})$
\EndFor
\EndProcedure
\end{algorithmic}
\caption{Chunking Method: AC in place}
\label{alg:chunk1}
\end{algorithm}

A variation of Algorithm~\ref{alg:chunk1}, {\sc kkmemGPUchunk2} given in Algorithm~\ref{alg:chunk2}, 
brings the row partition of $B$ 
into fast memory and streams through $A$ and $C$ chunks by switching
the two for loops with small changes.  Once it finds
$C^1_1 = A_1 \times B_1$, it moves to next chunks
$A_4$ and $C_2$ to perform $C^2_1 = A_4 \times B_1$. 
This variation copies $B$ 
onces, while $A$ and $C$ are copied as many times as the number of
partitions of $B$ ($C$ is partially copied in the first part).
The copy cost becomes 
($size(B) + size(A) \times \|P_{B}\| + size(C) \times (\|P_{B}\| -1) $).


\begin{algorithm}[t]
\smallerfont
\begin{algorithmic}[1]
\Procedure{{\sc kkmemGPUchunk2}}{$A,B,C, P_{AC}, P_{B}$}
\ForAll{$B_{rp} \in P_{B}$}
  \State $FB =copy2Fast(B, B_{rp})$ 
  \ForAll{$AC_{rp} \in P_{AC}$}
    \State $FC =copy2Fast(C, AC_{rp})$
    \State $FA =copy2Fast(A, AC_{rp})$
    \State $FC = kkmem(FA, FB, FC,AC_{rp}, B_{rp})$
    \State $copy2Slow(FC, C, AC_{rp})$
  \EndFor
\EndFor
\EndProcedure
\end{algorithmic}
\caption{Chunking Method: B in Place}
\label{alg:chunk2}
\end{algorithm}

Given the parts $P_{AC}$ and $P_{B}$, we calculate the data movement
cost of both methods, and we choose the method with the smaller cost. Note
that, the data movement of {\sc kkmemGPUchunk1} is minimized by reducing
the number of partitions of $A$ and $C$. This provides the maximum space
for $A$ and $C$ and we use the rest for $C$. Similarly, the copy cost
of the variation ({\sc kkmemGPUchunk2}) is minimized by reducing the number of partitions
of $B$, by providing $B$ the maximum space. However, when we follow
these approaches in practice, within a single multiplication, the
computation units will not be fully utilized. For this reason, we follow
the heuristic in Algorithm~\ref{alg:chunker} for determining the
partitions. Given the high-bandwidth memory size, we make sure that we
provide at least 25\% of the memory for the matrix that will be copied
in the inner loop. First, we try to place the whole matrix into fast
memory. If $B$ or $A$ and $C$ fits into the fast memory we achieve
optimal data movement cost. In this case, we leave the rest of the
memory to be used for the other matrices. With a binary search method, we
determine the ranges of the rows that fit into fast memory, and call the
chunk algorithm with the minimum cost. If neither $A$ and $C$, or $B$
fits into the fast memory, we check the data movement costs of $A$ and
$C$ with respect to $B$. We aim to minimize the data movement cost of
the larger one by choosing the algorithm that copies it in the outer
loop. We further aim to minimize the cost of the inner loop by giving
the larger cost matrix the big portion of the fast memory so that its
number of partitions are minimized. Once the number of partitions are
calculated we call the algorithm with lower copy cost.

\begin{algorithm}[t]
\smallerfont
\begin{algorithmic}[1]
\Procedure{Partition}{$A,B,C, FastSize$}
  \State $BigPortion = 0.75 \times FastSize$
  \State $SmallPortion = 0.25 \times FastSize$
  \If{$size(B) < BigPortion$}
     \State $P_B = [(0, n)]$ {\it //n  = num rows of B}
     \State Add left over from big to small portion 
     \State Find the balanced partition size $pSize_{AC}$ for $A$ and $C$
     \State $P_{AC} = BinarySearch(A,C,pSize_{AC})$
     \State $kkmemGPUchunk2(A,B,C,P_{AC}, P_{B})$
  \ElsIf {$size(A) + size(C) < BigPortion$}
     \State Same as above but $A$ and $C$ gets bigger portion
     \State $kkmemGPUchunk1(A,B,C,P_{AC}, P_{B})$
  \ElsIf {$size(A) + 2 \times size(C) > size(B)$}
     \State Find $pSize_{AC}$ for $A$ and $C$ using big portion 

     \State $P_{AC} = BinarySearch(A,C,pSize_{AC})$
     \State Add left over from big to small portion 
     \State Find the balanced partition size $pSize_{B}$ for $B$ 
     \State $P_{B} = BinarySearch(B,pSize_B)$
     \State choose the heuristic with lower copy cost
  \Else
     \State Same as above where B gets the larger portion
     \State choose the heuristic with lower copy cost
  \EndIf
\EndProcedure
\end{algorithmic}
\caption{Chunking Decision Heuristic}
\label{alg:chunker}
\end{algorithm}


\myspace{-0.5ex}
\section{Evaluation of Data Placement and Chunking Methods} 
\label{sec:results}
\myspace{-0.5ex}

\subsection{KNL Experiments}

\noindent
This section evaluates the performance of the selective data placement technique (DP) 
and the chunking method for KNLs. Comparison of the baseline method {\sc{kkmem}}
against state-of-art \spgemm{} literature can be found 
in~\cite{deveci2017performance,deveci2018multi}, 
and is out of the scope of this paper, which focuses on the evaluation 
of algorithm performance on multilevel-memory systems.

\subsubsection{Multigrid Compututations} 

\begin{figure*}
\begin{center}
\subfloat[Laplace]{\includegraphics[width=\imscale\columnwidth]{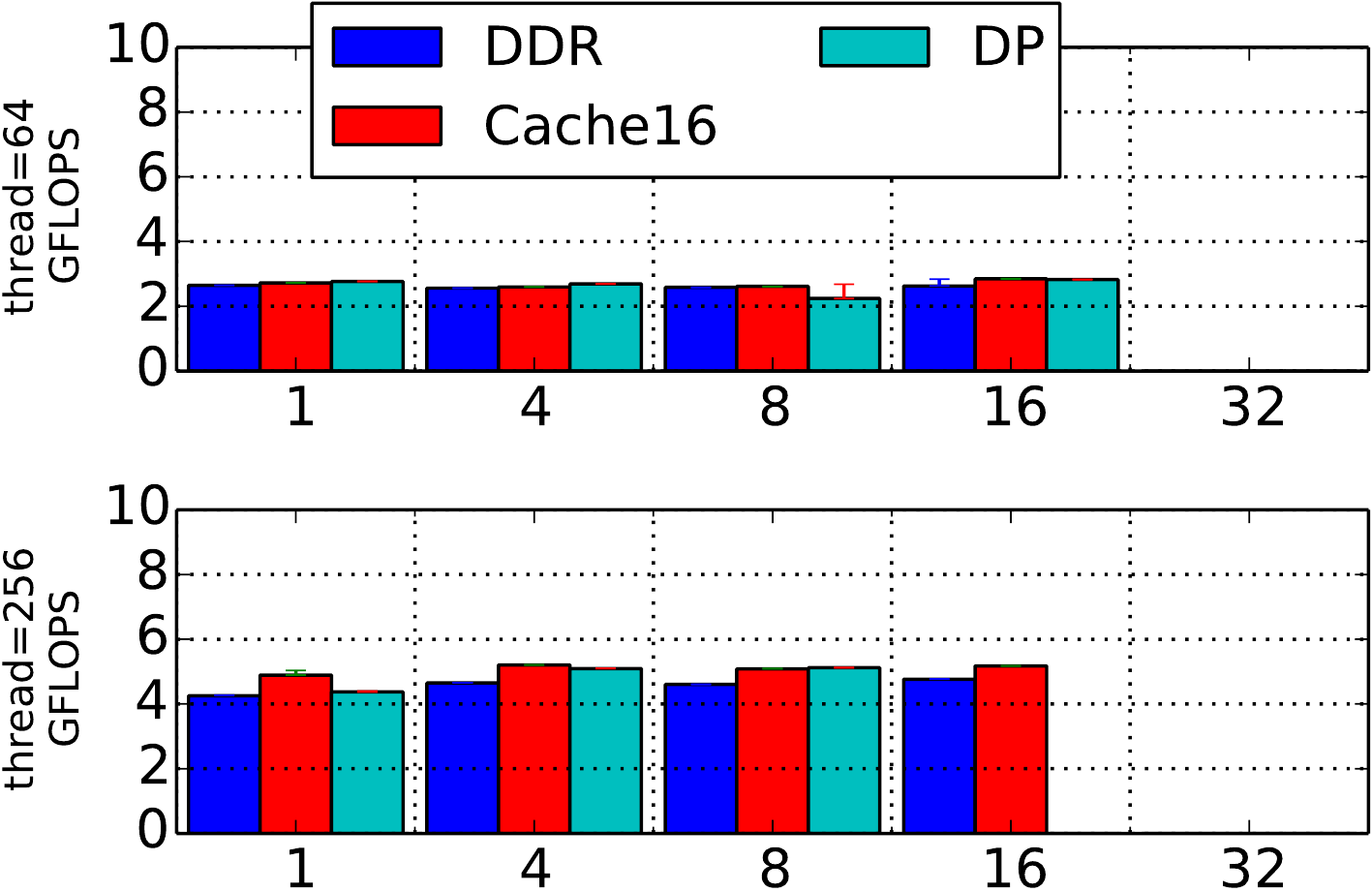}\label{fig:explapap}}
\subfloat[BigStar]{\includegraphics[width=\imscale\columnwidth]{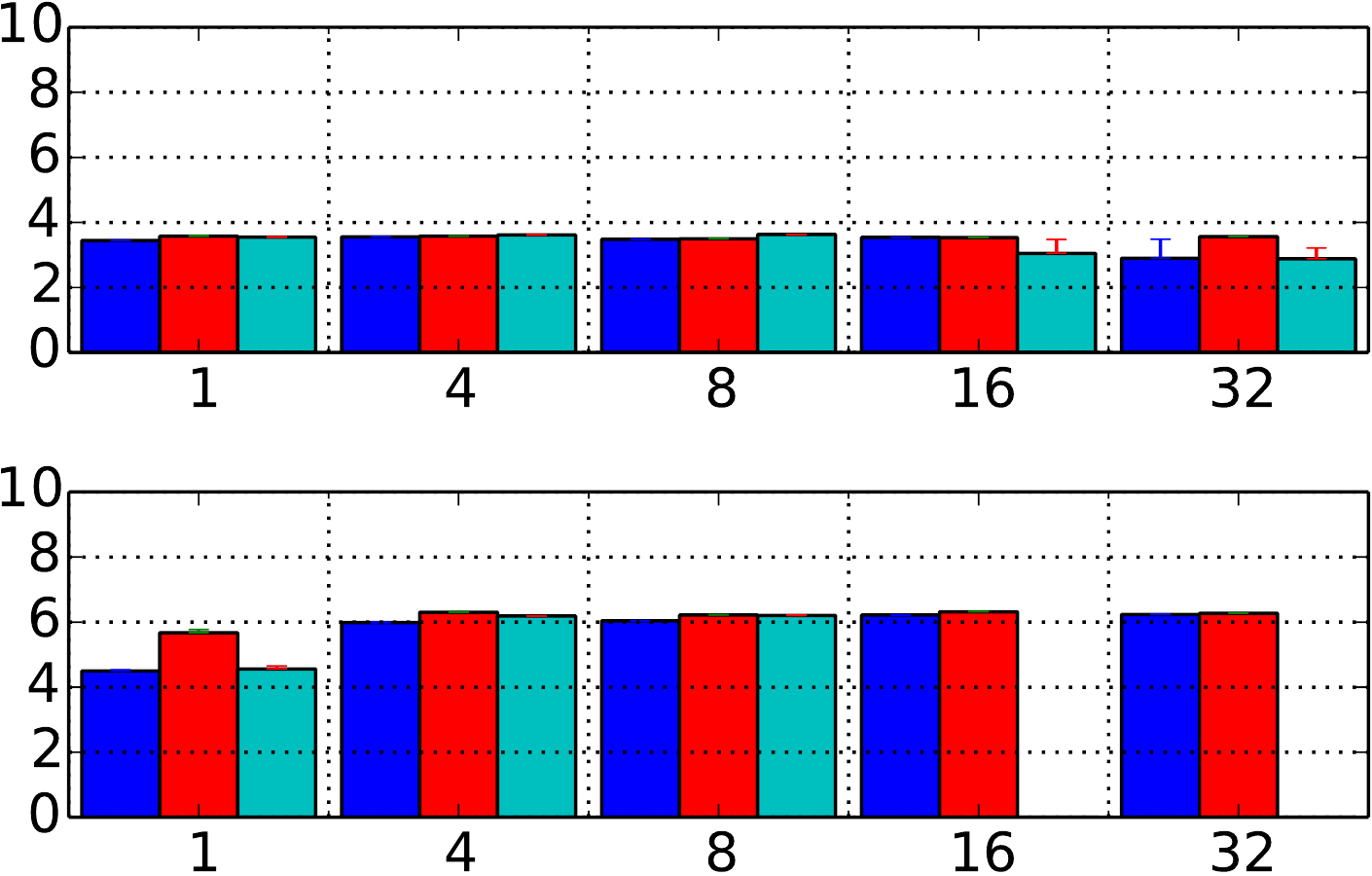}\label{fig:expbigap}}

\subfloat[Brick]{\includegraphics[width=\imscale\columnwidth]{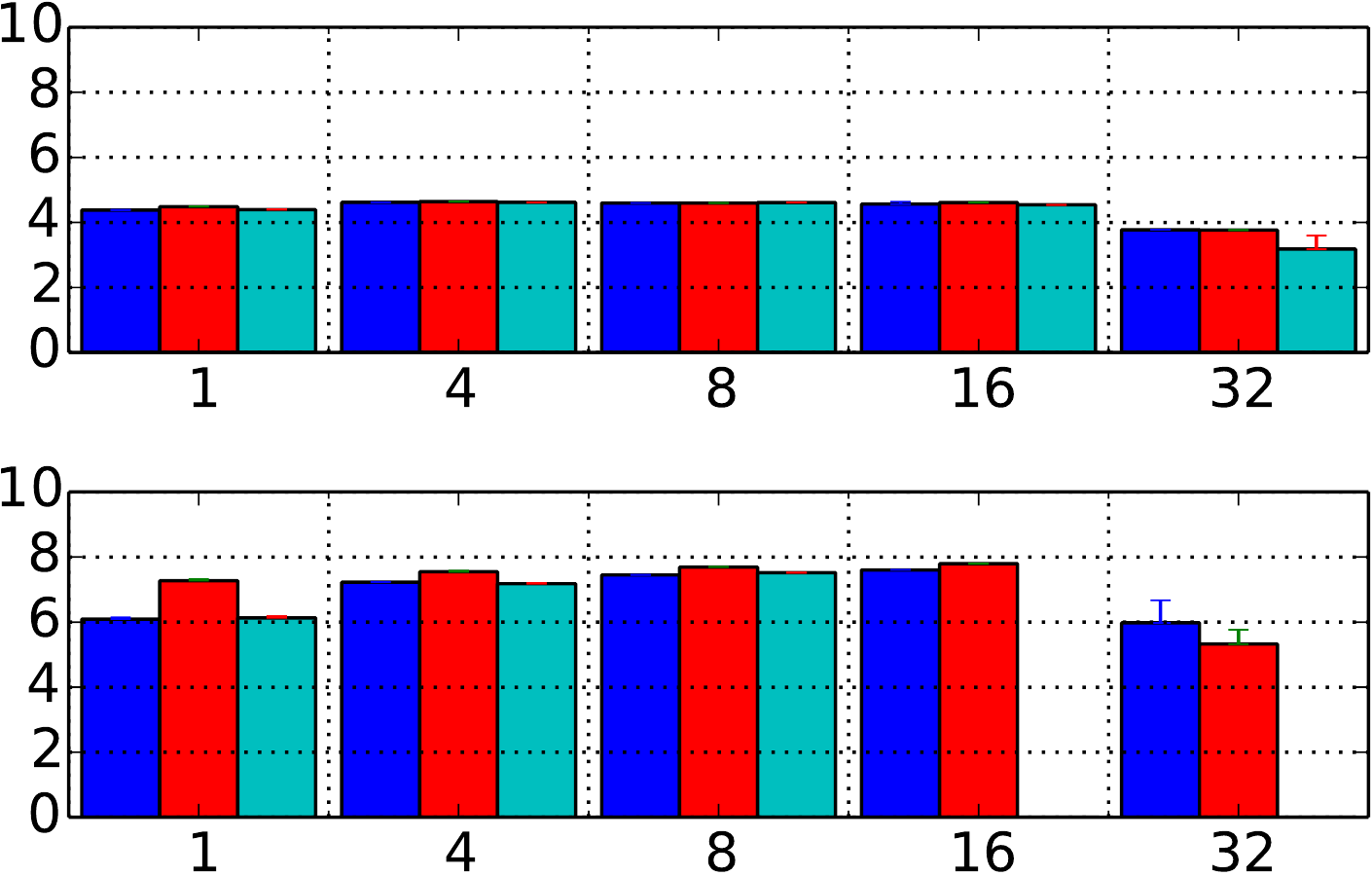}\label{fig:expbrickap}}
\subfloat[Elasticity]{\includegraphics[width=\imscale\columnwidth]{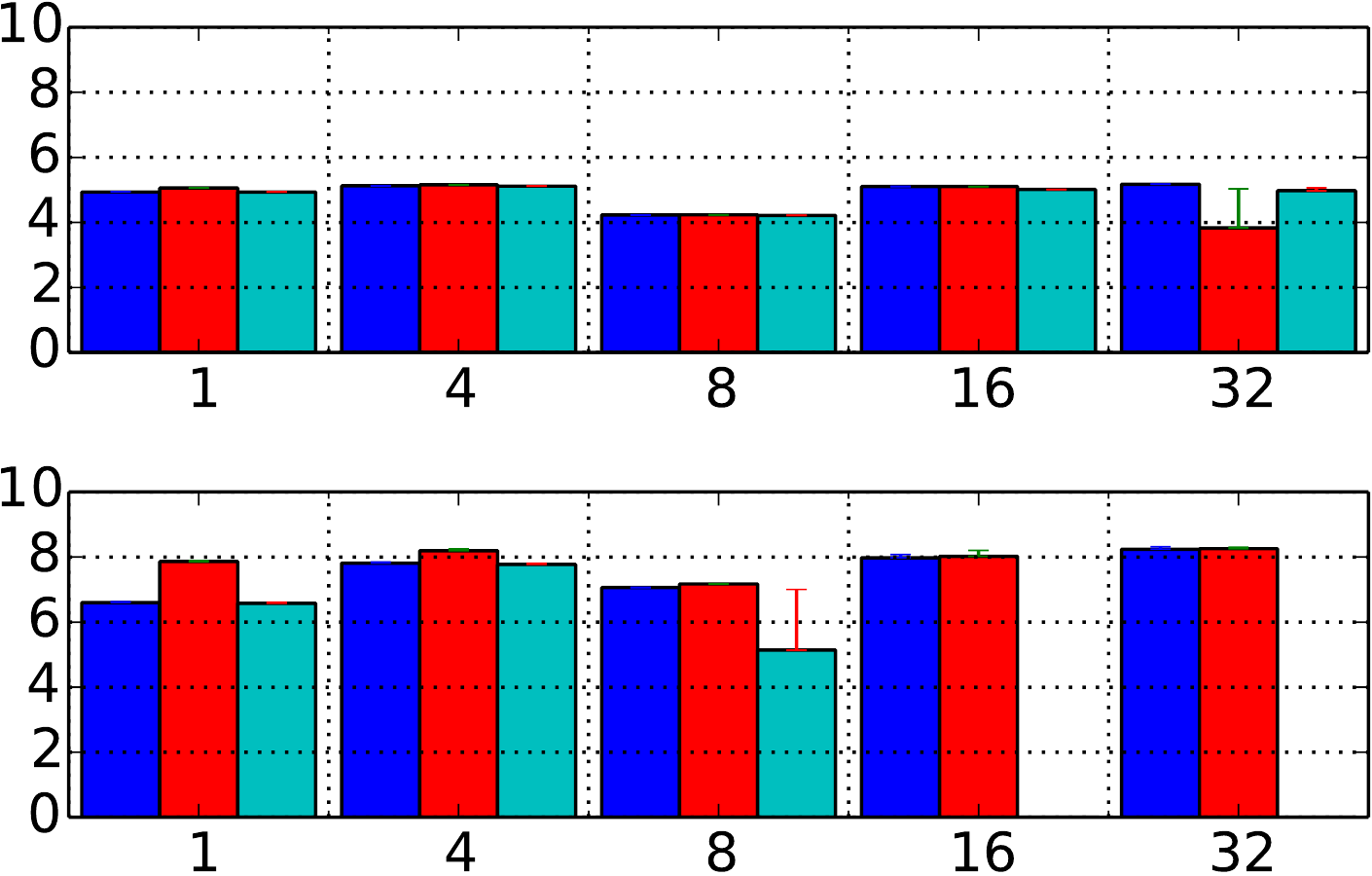}\label{fig:expelasap}}

\end{center}
\caption{$A\times P$ Multiplications on KNL. }
\label{fig:expknlap}
\end{figure*}

\begin{figure*}
\begin{center}

\subfloat[Laplace]{\includegraphics[width=\imscale\columnwidth]{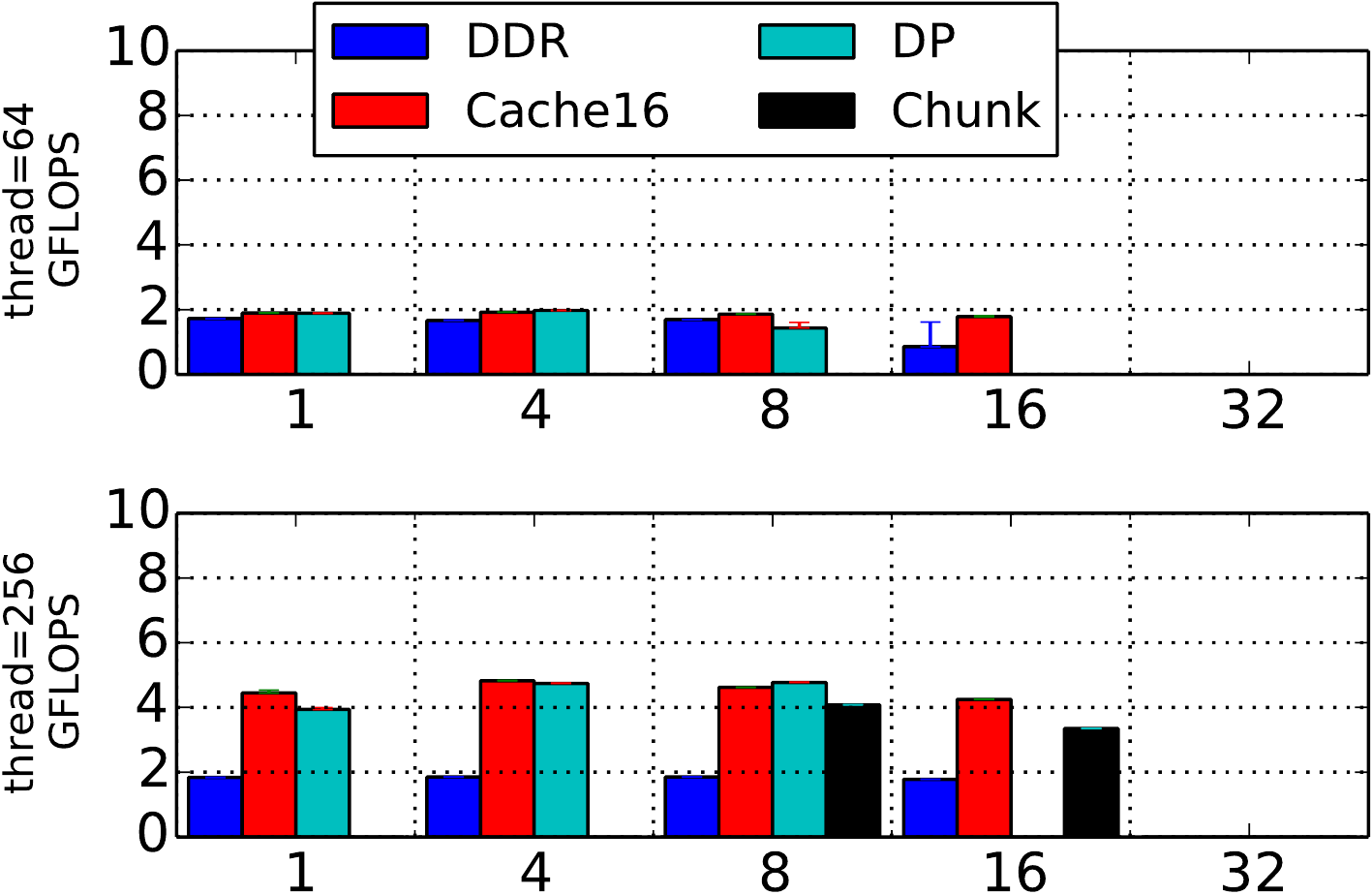}\label{fig:explapra}}
\subfloat[BigStar]{\includegraphics[width=\imscale\columnwidth]{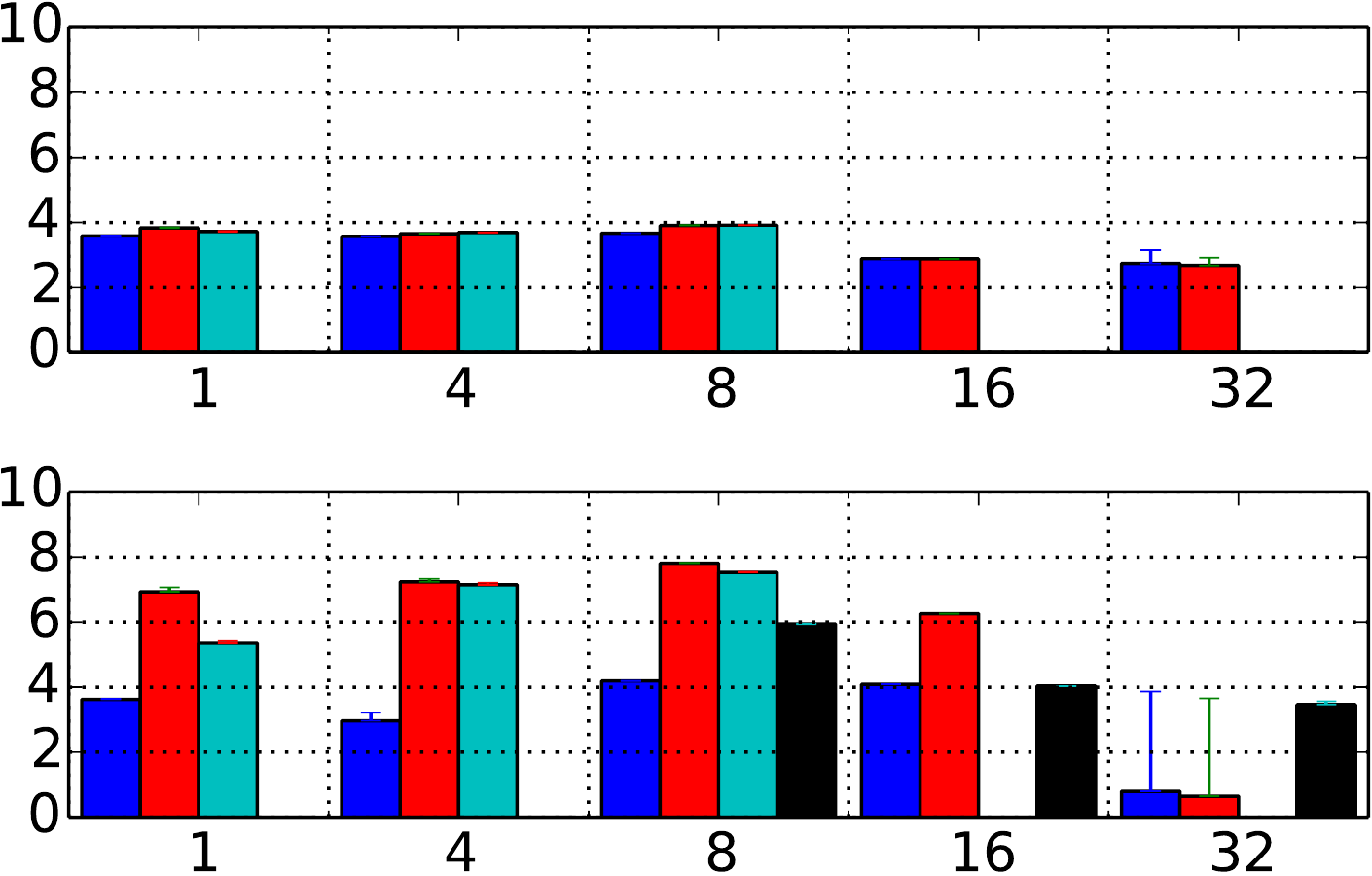}\label{fig:expbigra}}

\subfloat[Brick]{\includegraphics[width=\imscale\columnwidth]{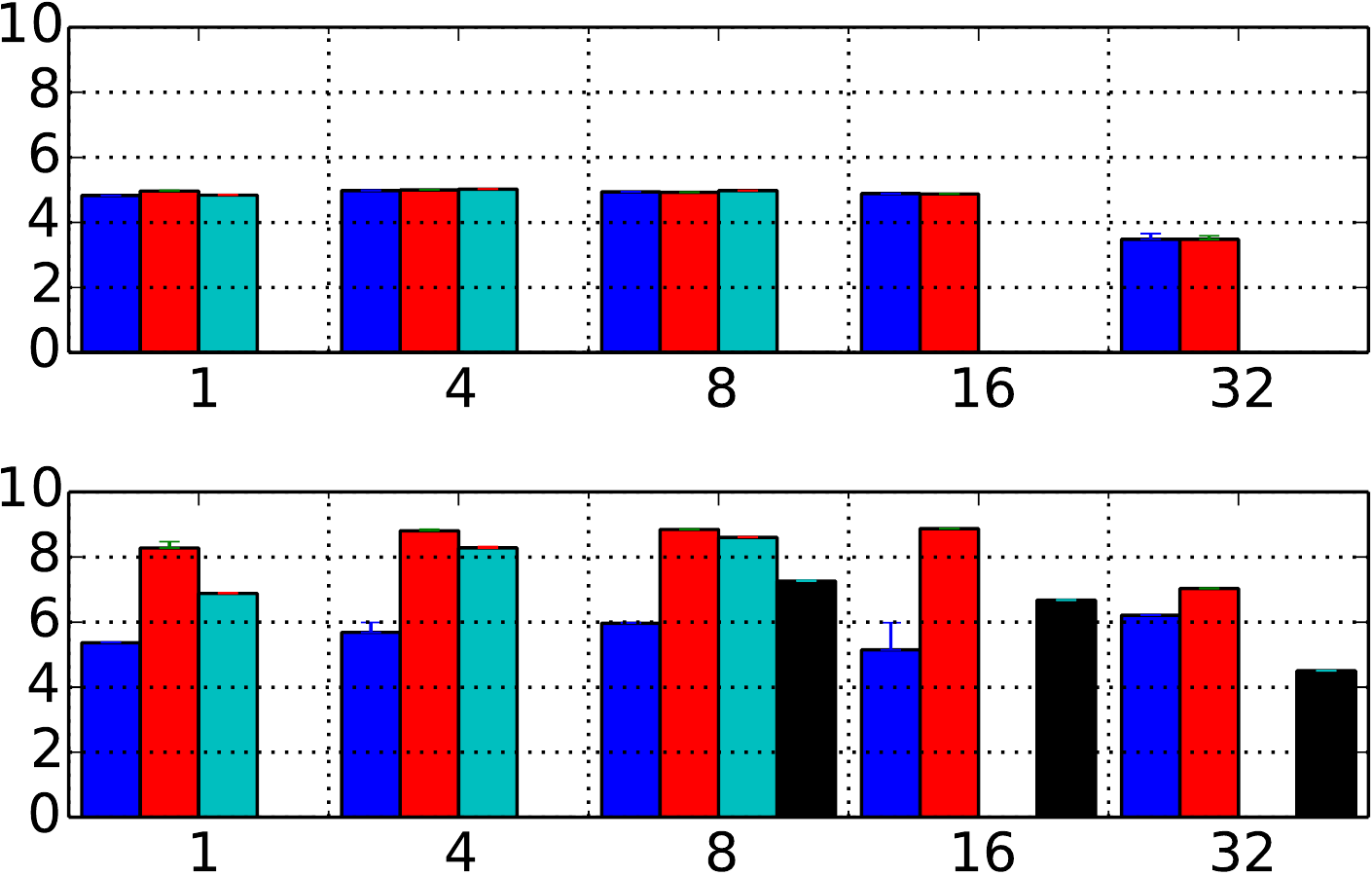}\label{fig:expbrickra}}
\subfloat[Elasticity]{\includegraphics[width=\imscale\columnwidth]{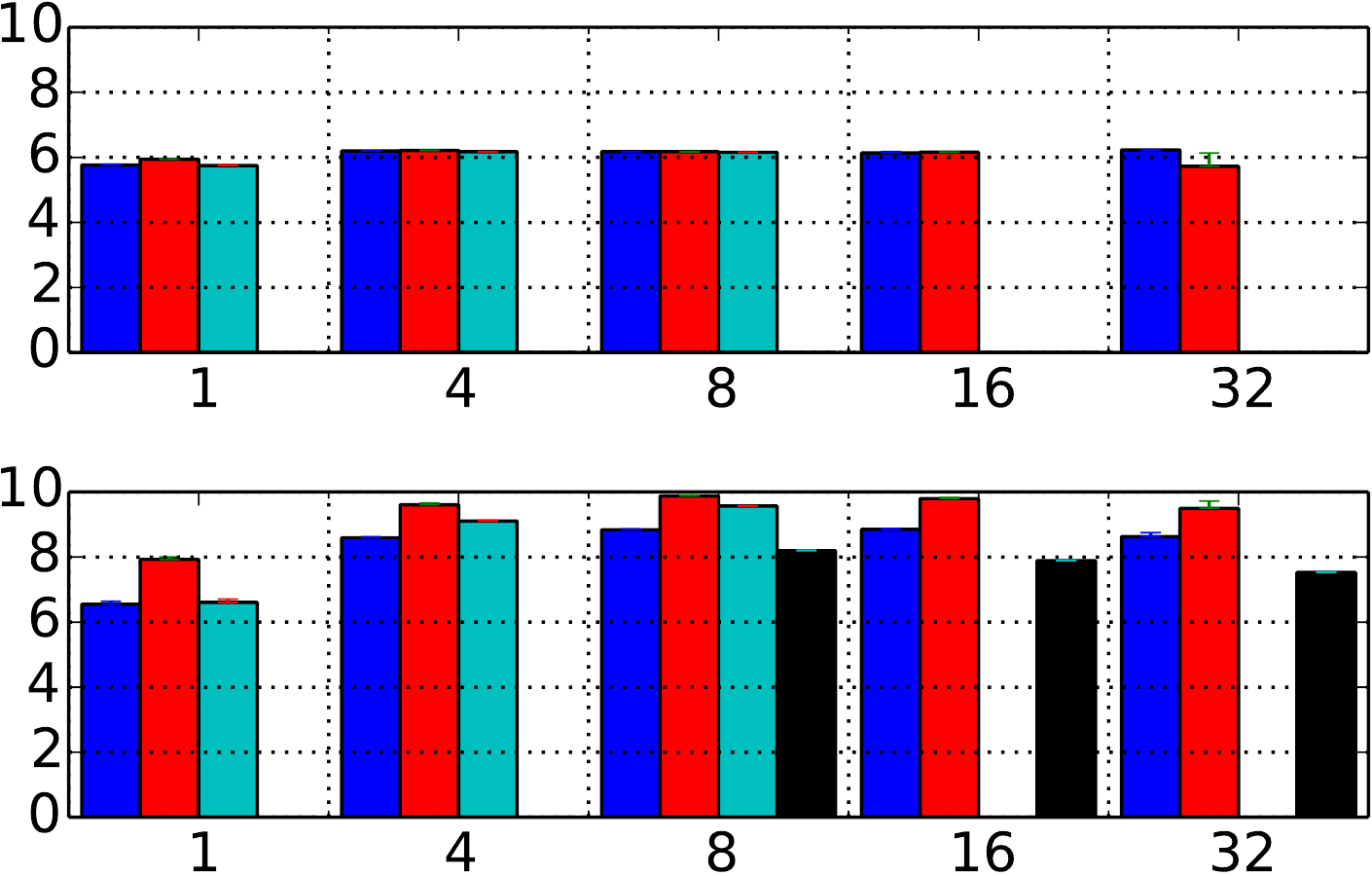}\label{fig:expelasra}}
\end{center}
\caption{$R \times A$ Multiplications on KNL. }
\label{fig:expknlra}

\end{figure*}

\noindent
Figure~\ref{fig:expknlap} and~\ref{fig:expknlra} present the GFLOP/s achieved 
on $A\times P=AP$ and $R\times A=RA$ multiplications. We only show Cache16 and DDR 
from Figure~\ref{fig:knlap} and~\ref{fig:knlra} for the simplicity of figures.
For all $A\times P$ multiplications, $P$ is smaller than the HBM size,
thus, allowing us to run DP for all instances. The performance of {\sc{
kkmem}} across different memory systems is similar in these problems.
This is also the case for DP method as well. 
We do not observe a significant benefit of HBM with DP; because
of the regular structure of $A$, the accesses to $P$ are not expensive.
Moreover, the sizes of $A$ and $AP$ are much larger than $P$. As a
result, the accesses to $P$ are not the dominant memory operations in
these multiplication operations.

For $R\times A$ multiplications, $A$ has both the
largest size, and irregular access patterns because of the
structure of $R$. As a result, the accesses to $A$ are the most dominant
memory operations. DP significantly benefits from placing $A$ into HBM.
In most cases, placing $A$ on HBM alone recovers the performance
drop of using DDR as the main area for allocation, and makes it very
close to the performance of HBM (or Cache16). However, DP only works when $A$ fits into HBM.

We run the chunking algorithm for KNLs only for $R\times A$ multiplication for
256 threads. We expect a performance improvement only for this case.
For other cases, the performances of HBM and DDR are similar, and the data
movement cost introduced by chunking further reduces the
performance. We run the chunked algorithm where fast memory size is limited to
8GB (as allocations exceeding 11GB led to exhaustion of
the memory capacity due to memory fragmentation). For the
inputs where $A$ is 8GB, the algorithm copies all of $A$ into HBM and performs
the multiplication. The performance of the core multiplication kernel 
is same as DP; however, the cost of the data movement drops the
overall GFLOP/s achieved by 10\%. However, for bandwidth bound
$R\times A$ multiplications, this copy can still improve the performance w.r.t. DDR (except
Elasticity). For larger inputs, the copy overhead only amortizes for
those that greatly benefit from HBM. However, when the multiplication performance 
is similar on HBM and DDR, as in Elasticity,
the use of chunking introduces a copy overhead and drops the performance.

In conclusion, DP and chunking benefit SpGEMM on KNLs, where memory
systems differ in terms of bandwidth, only when:
\begin{enumerate}
\item Accesses to $B$ are irregular (lower temporal locality).
\item A row of $B$ is accessed many times without temporal locality.
\item The density of the rows in $B$ is small (lower spatial locality).
\end{enumerate}

Point (1) holds for $R\times A$ multiplications. However, each column
appears around 3 to 4.5 times in these problems, breaking the second
condition. Such lower reuse tends to not amortize the copy cost. (3) varies
over the matrices, and it is lowest for $R\times A$ of Laplace problem.
We get the most benefit in this problem. $A\times P$ has many accesses
to $P$ but these accesses show temporal locality.

\subsubsection{Triangle Counting}

\begin{figure}
\begin{center}
\subfloat[twitter]{\includegraphics[width=\imtscale\columnwidth]{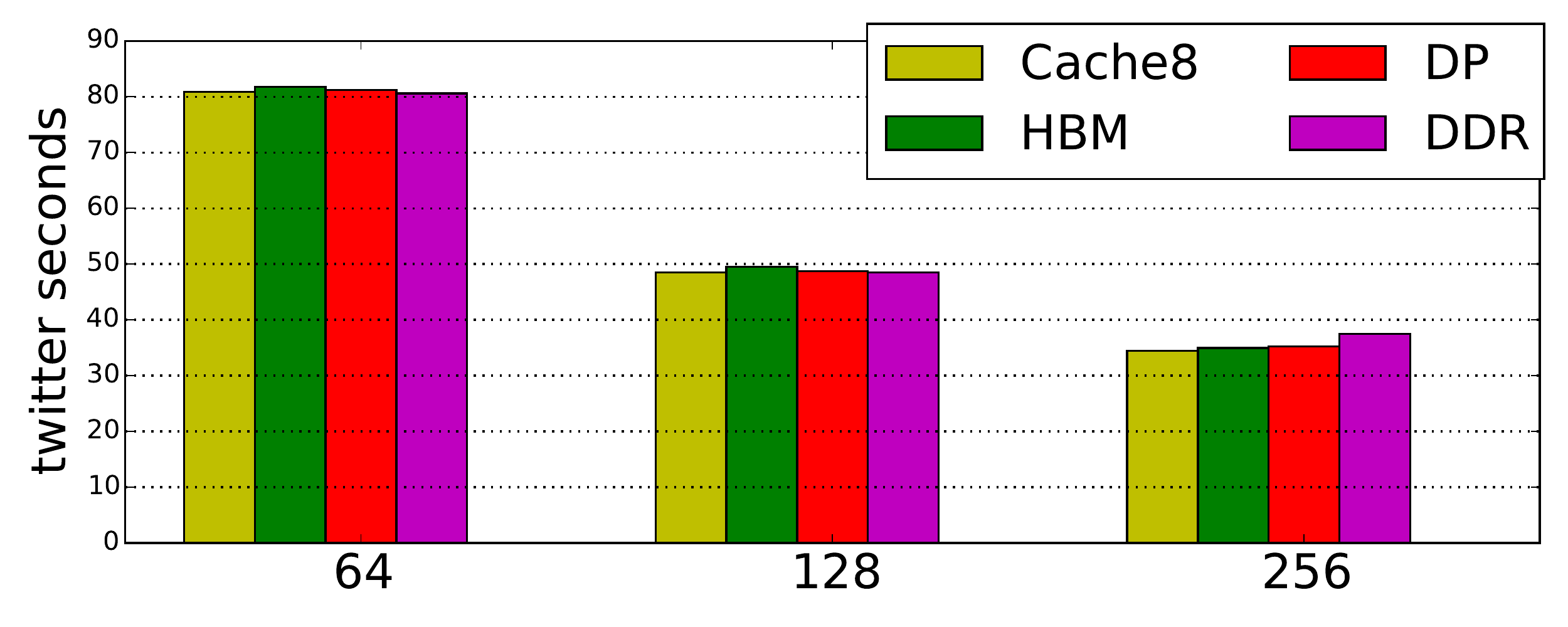}}

\subfloat[uk\_2005]{\includegraphics[width=\imtscale\columnwidth]{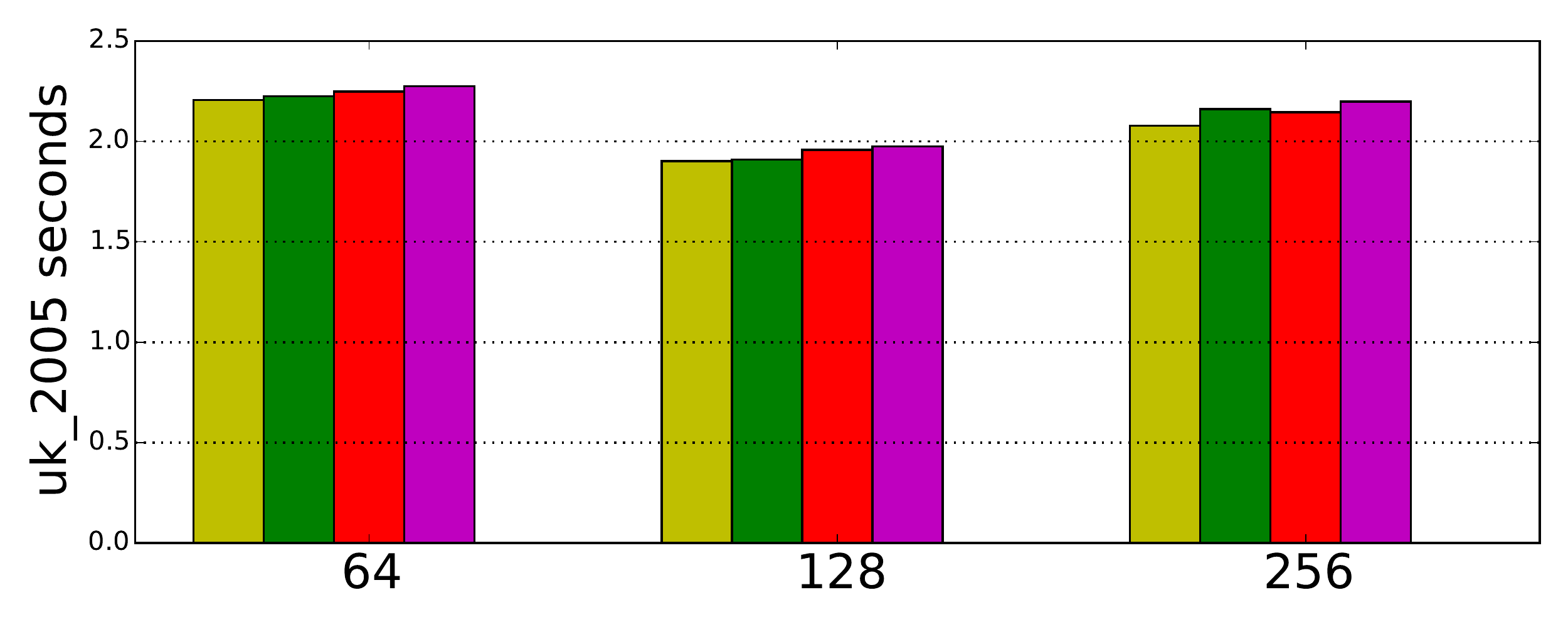}}

\subfloat[g500s25f16]{\includegraphics[width=\imtscale\columnwidth]{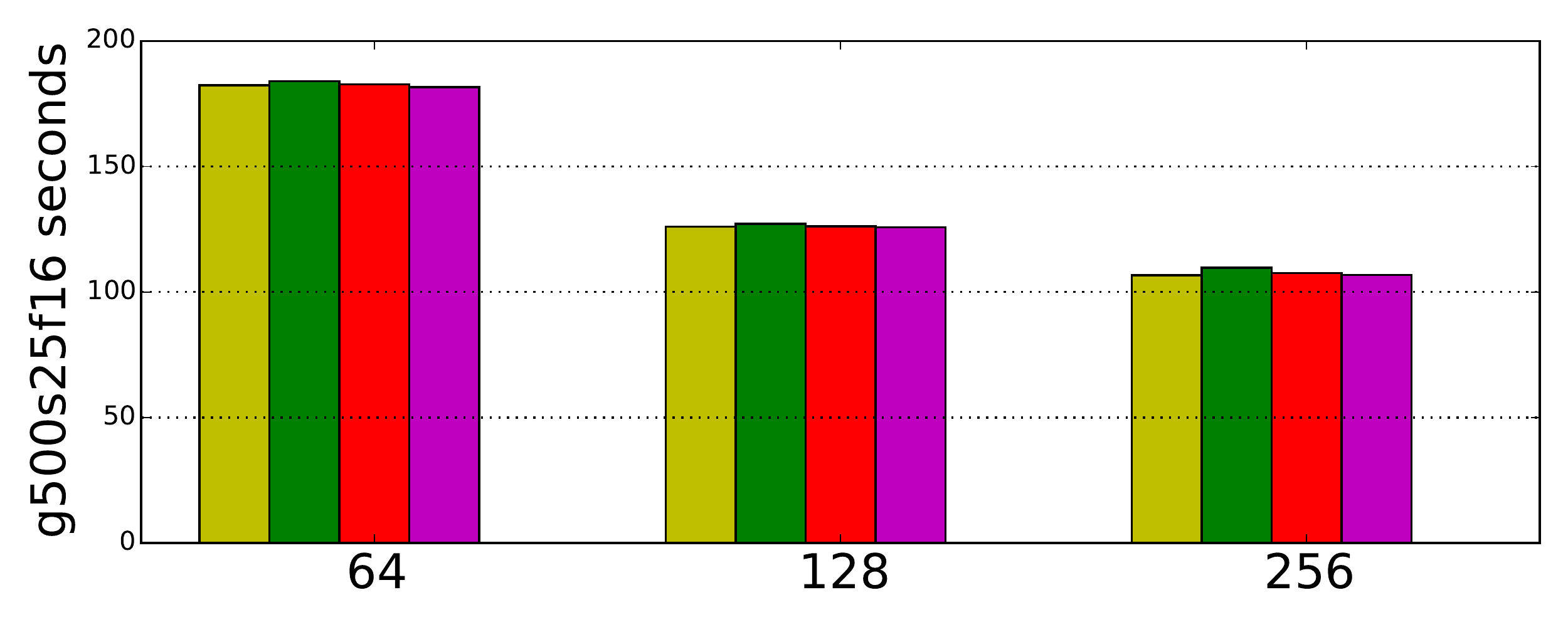}}
\end{center}
\caption{Triangle counting times in seconds.}
\label{fig:exptriangle}
\end{figure}

So far, we have exclusively used matrices from multigrid
solvers. This was mainly because it allows us to perform weak scaling
studies and evaluate two extreme access patterns: regular $A\times P$,
highly irregular $R\times A$. In this section, we demonstrate
characteristics on the application of \spgemm{} when used to perform
triangle-counting -- a graph analytics problem of interest.

Triangle counting problem is used in many network analysis applications,
including social network analysis, spam detection, link recommendation,
and dense neighborhood discovery. For these experiments we use
three graphs from these applications areas to replicate realistic use
cases. These graphs include: (1) twitter-2010, a social network graph; (2)
uk-2005, web crawl graph, and, finally, (3) g500s25f16, which is a
graph500 scale25 graph. In these experiments, we use an extension of {\sc kkmem}
proposed by Wolf {\em et al.}~\cite{wolf2017fast} for triangle counting. 
In this method, the rows are sorted in terms of their degree and a 
lower triangular matrix is multiplied and masked with itself. The sum of the
overall values correspond to the number of triangles in the graph. 
Wolf {\em et al.} uses {\sc kkmem} with a fused masking
operation for this triangle counting. 
We use the same technique with data placement using {\sc kkmem} as all matrices 
fit into HBM.

In the triangle counting problem we work only on the symbolic structure
of matrix - there is no output matrix. Since {\sc kkmem} performs a
symbolic compression of the right hand side matrix, our matrix
multiplication kernel simply computes $L \times compressed(L)$. 
In DP, we only place compressed $L$ into HBM. In
Figure~\ref{fig:exptriangle}, we give the overall runtime of the
triangle counting for three matrices in DDR.
These times include preprocessing
times such as sorting, creation of lower triangular matrices, as well as
 compression time. However matrix multiplication time dominates (using up
to 95\% of execution time except uk\_2005 which is 30\%).
In this experiment, 
all memory modes obtain similar performances on the same number of 
threads. The triangle counting kernel is also oblivious to the 
bandwidth characteristics of the underlying memory system. 
Among these matrices, the kernel scales well using all
hyperthreads on twitter and g500s25f16. Table~\ref{tab:exptricache}
shows the profiled L1 and L2 cache-miss ratios provided by Kokkos~\cite{edwards2014kokkos}
Profiling tools. As shown, uk\_2005 has the highest percentage of cache
misses on L2. The more frequent access to the memory system will drive
the problem to become bandwidth bound more quickly which is also the
reasoning for the scaling issues seen with 256 threads.
In these problems, the right hand side has denser rows, while left hand side has
irregular access patterns. As a result,
these problems are similar to $R\times A$ multiplication of the Elasticity
problem. The performance trends also show similarities where memory
system or data placement has minimal effect on the performance. 
As a result, a chunking method with extra copy cost is not critical and it is 
likely to reduce the performance in these problems.

\begin{table}[]
\centering
\caption{Cache miss rates for triangle counting on KNL for 64 threads.}
\label{tab:exptricache}
\begin{tabular}{c|r|r}
\multicolumn{1}{l|}{} &  \multicolumn{1}{c|}{L1-M\%} & \multicolumn{1}{c}{L2-M\%} \\ \hline
\multicolumn{1}{c|}{g500s25f16} &  0.78 & 4.63 \\ \hline
\multicolumn{1}{c|}{twitter} & 0.24 & 16.95 \\ \hline
\multicolumn{1}{c|}{uk-2005} & 0.09 & 18.19 \\ 
\end{tabular}
\end{table}

\subsection{GPU Experiments} 

\begin{figure*}
\begin{center}
\subfloat[Laplace]{\includegraphics[width=\imtscale\columnwidth]{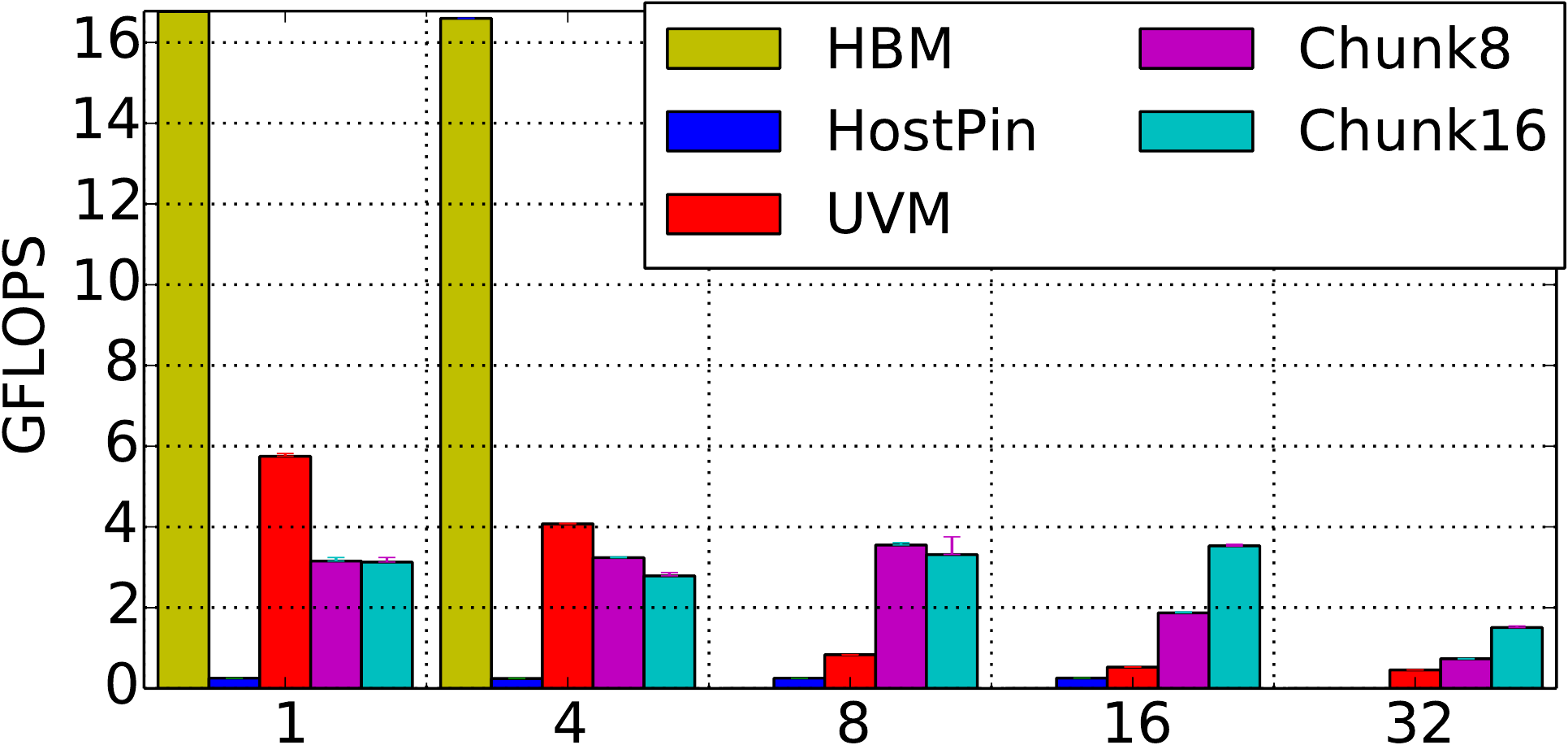}\label{fig:expgpulapap}}
\subfloat[BigStar]{\includegraphics[width=\imtscale\columnwidth]{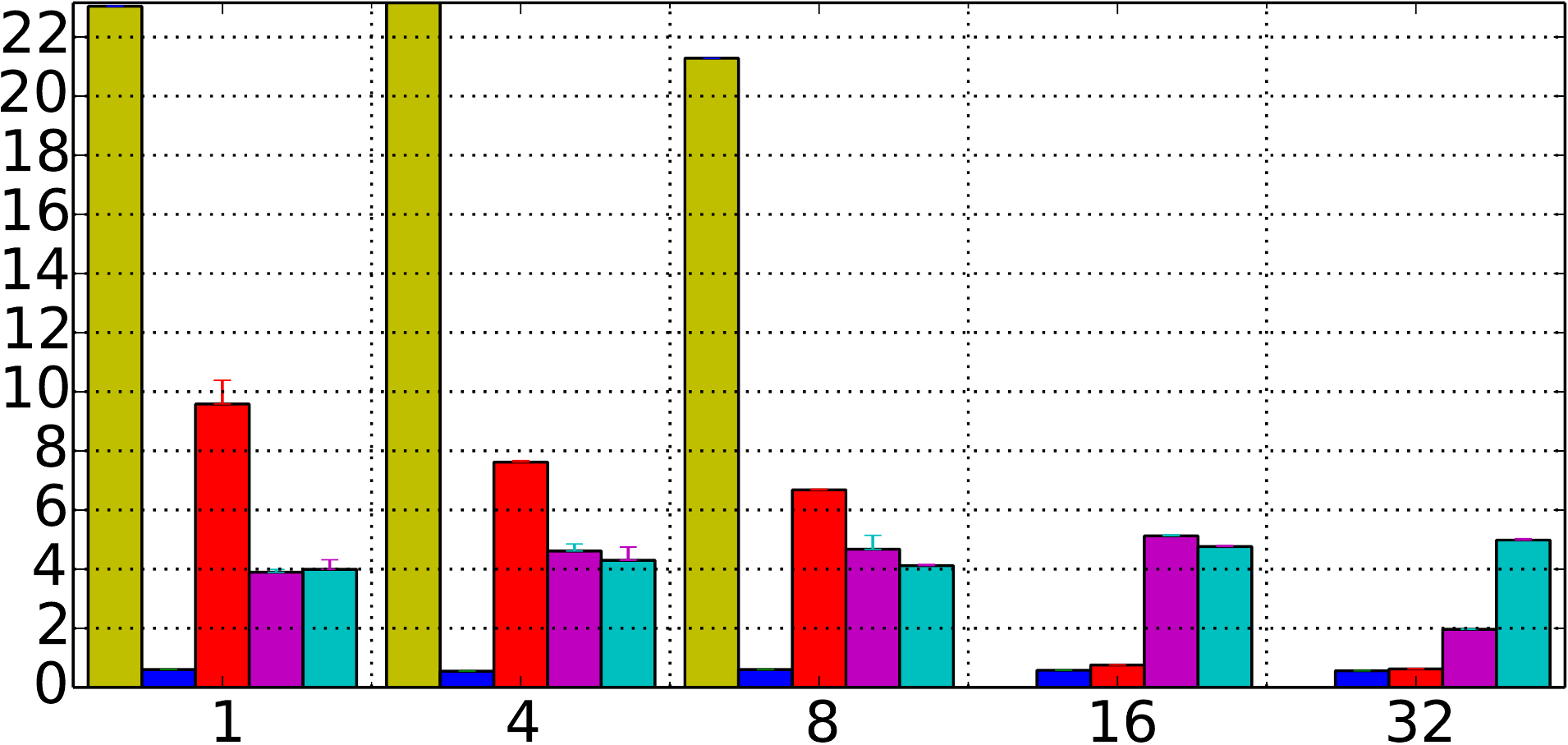}\label{fig:expgpubigstarap}}

\subfloat[Brick]{\includegraphics[width=\imtscale\columnwidth]{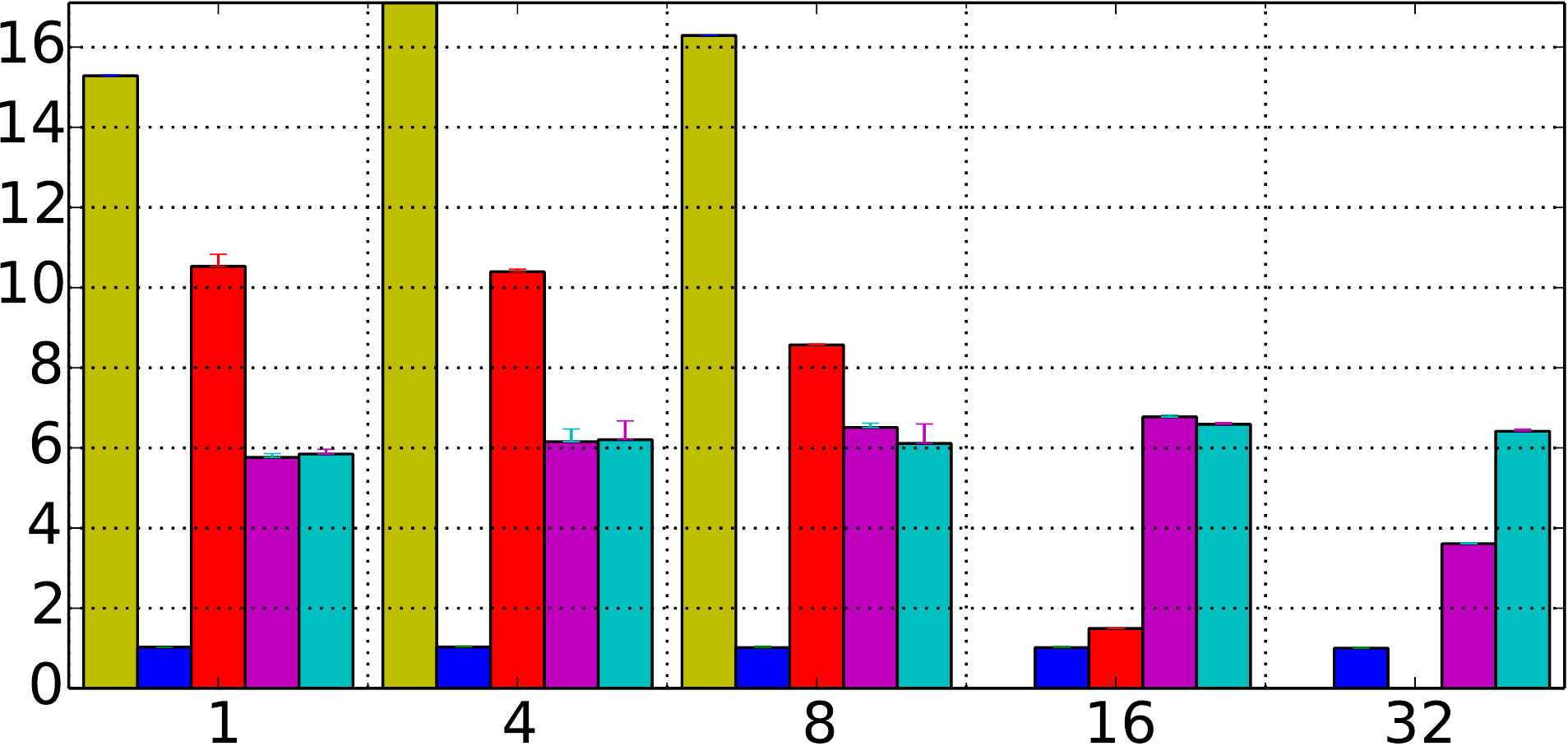}\label{fig:expgpubrickap}}
\subfloat[Elasticity]{\includegraphics[width=\imtscale\columnwidth]{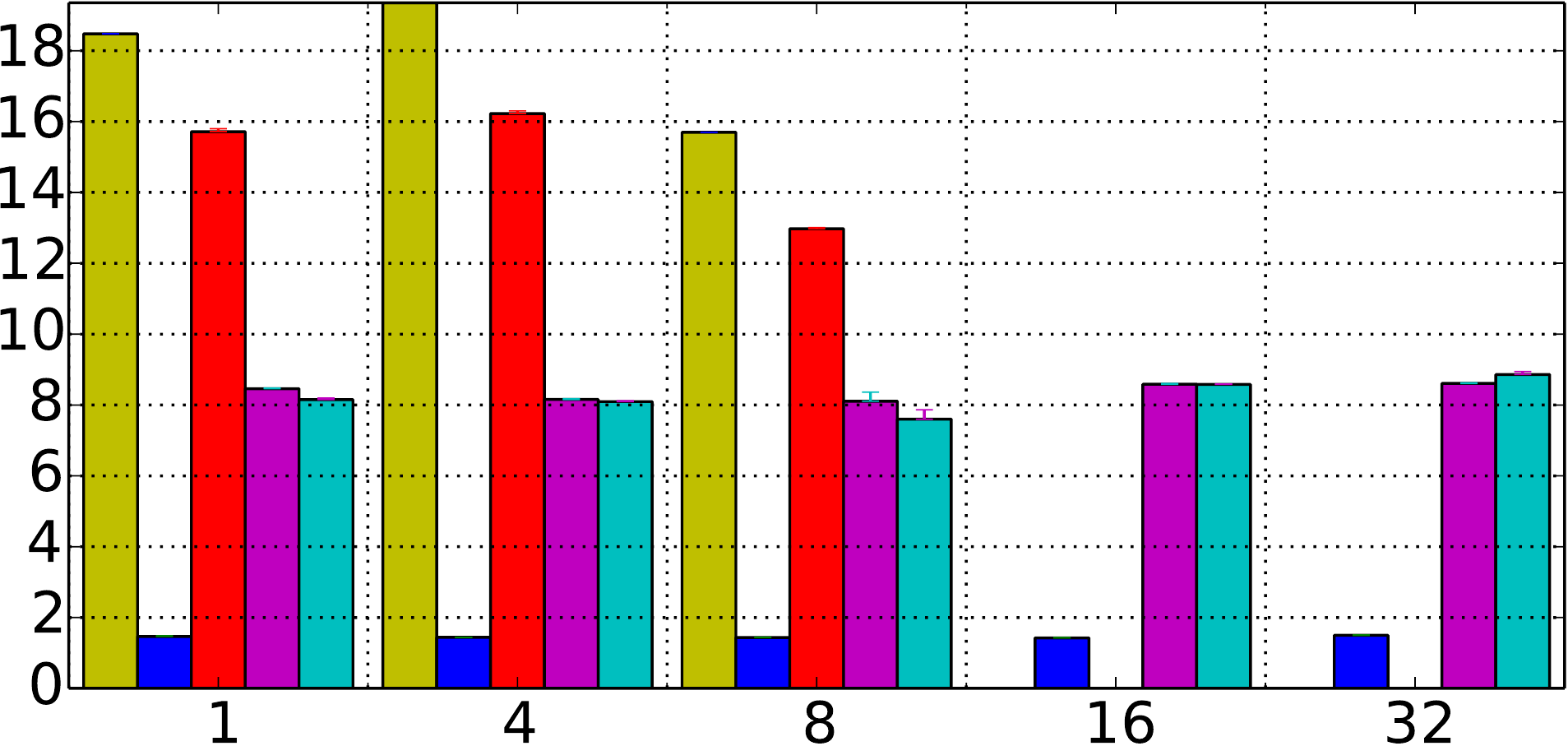}\label{fig:expgpuelasticityap}}
\end{center}
\caption{Algorithmic GFLOP/s achieved by HBM, Pinned Memory, and UVM and Chunked Algorithms on $A \times P$ multiplications.}
\label{fig:expgpuap}
\end{figure*}

\begin{figure*}
\begin{center}
\subfloat[Laplace]{\includegraphics[width=\imtscale\columnwidth]{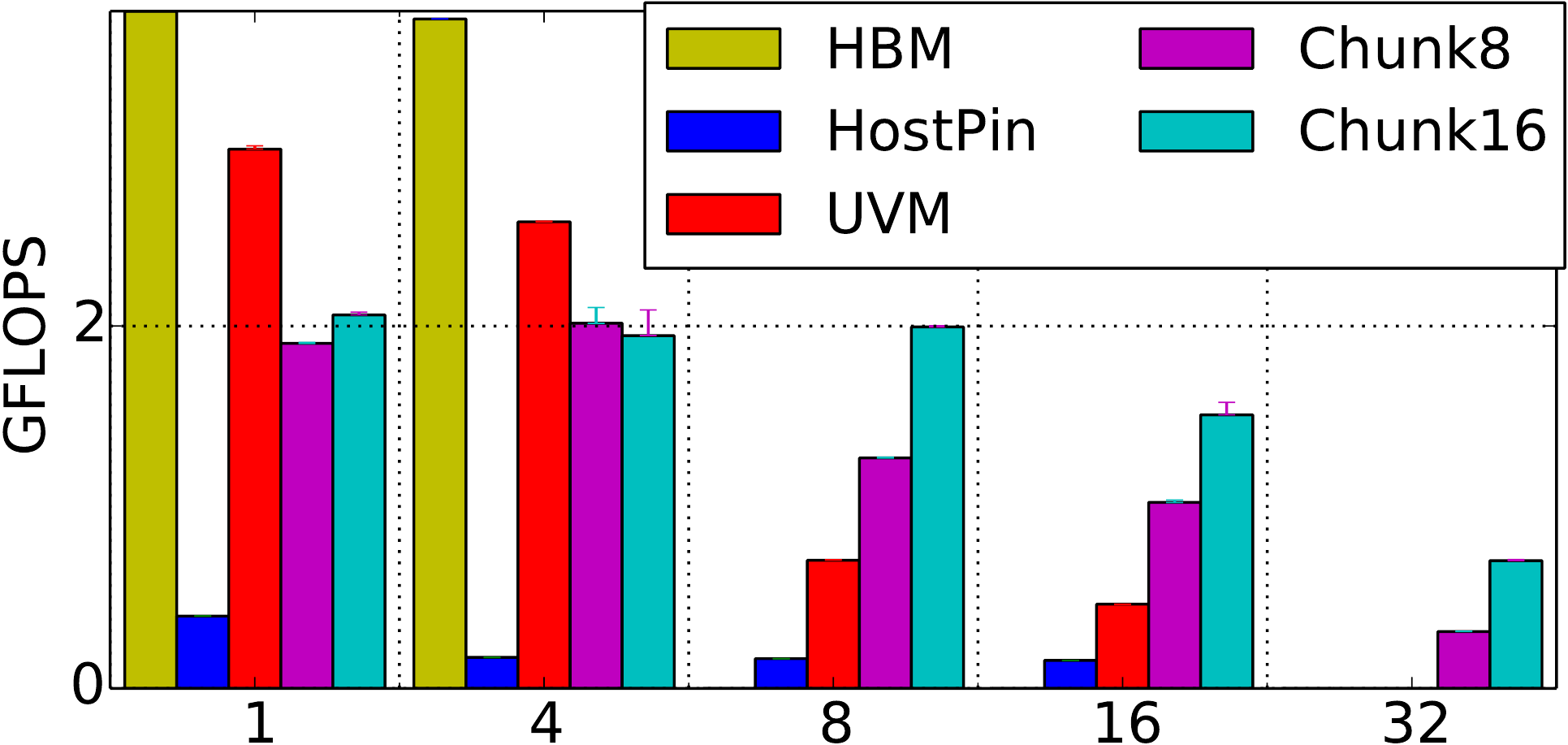}\label{fig:expgpulapra}}
\subfloat[BigStar]{\includegraphics[width=\imtscale\columnwidth]{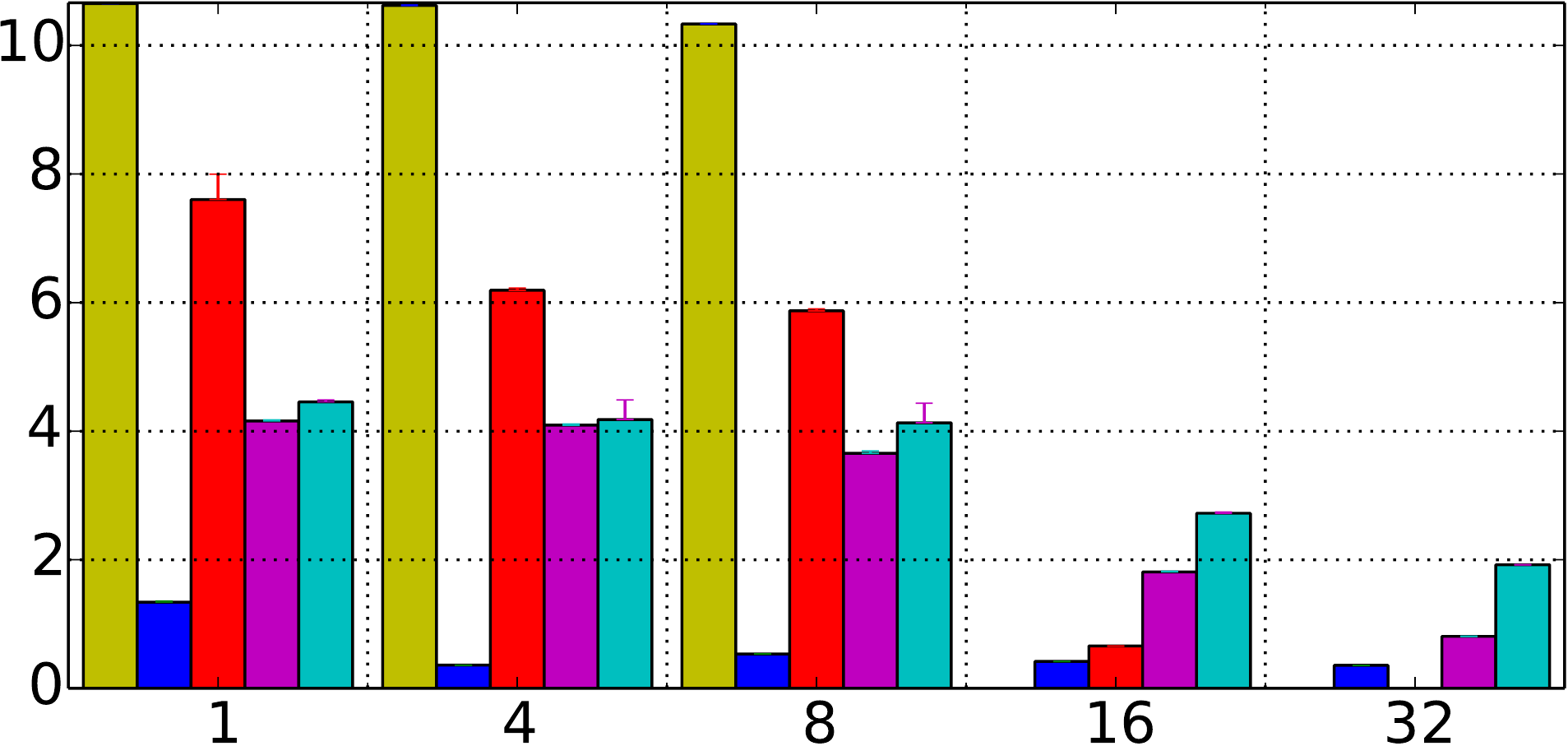}\label{fig:expgpubigstarra}}

\subfloat[Brick]{\includegraphics[width=\imtscale\columnwidth]{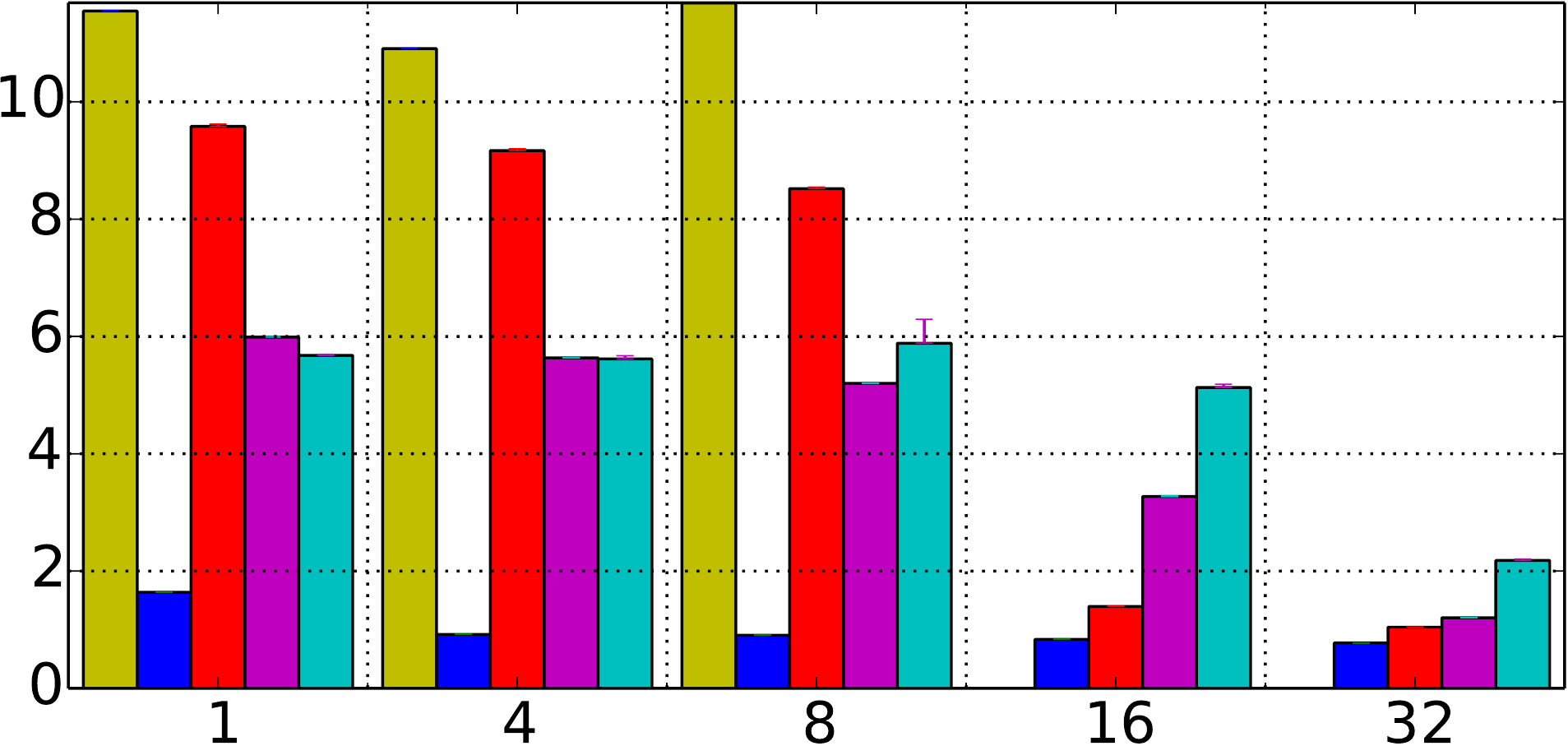}\label{fig:expgpubrickra}}
\subfloat[Elasticity]{\includegraphics[width=\imtscale\columnwidth]{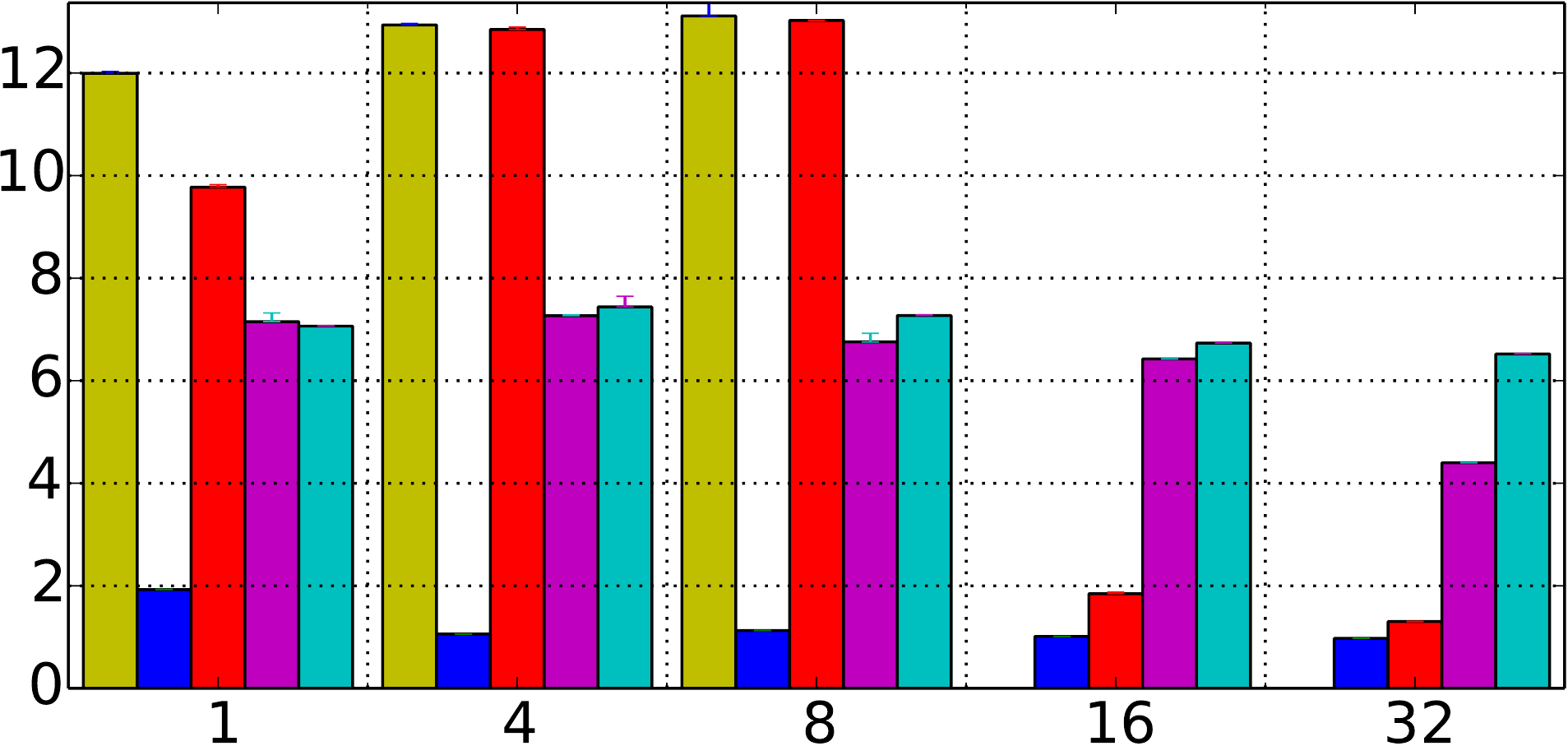}\label{fig:expgpuelasticityra}}
\end{center}
\caption{ Algorithmic GFLOP/s achieved by HBM, Pinned Memory, and UVM and Chunked Algorithms on $R \times A$ multiplications.}
\label{fig:expgpura}
\end{figure*}

\noindent
In this section we evaluate the performance of the proposed chunking
method on GPUs. Figures~\ref{fig:expgpuap} and~\ref{fig:expgpura} follow the same structure as in
Figure~\ref{fig:gpuap} and~\ref{fig:gpura}, and includes the results for two chunking
options. Chunk8 and Chunk16 run chunking operations where the
fast memory sizes are limited to 8 and 16GBs, respectively.

For problems where $A$ is smaller than 4GB, both Chunk8 and Chunk16 fit
whole problems into fast memory. In this case, there is no chunking that
is performed. Instead, the whole problem is copied into the fast memory.
Multiplication is performed in the fast memory and the result is copied
back to the host. In these cases, the multiplication kernel achieves the
performance of HBM; however the data movement costs reduce the achieved
FLOP/s. We observe at most 5.7$\times$ performance drop w.r.t. HBM
performance, yet we achieve as high as 14.7$\times$ speedup w.r.t. host
pinned memory runs. UVM obtains better performance than chunking when
the problem fits into HBM. Note that in these cases the data is already
in GPU, and there is no copy out to host. As a result, UVM should have no
data movement costs. UVM performance drops significantly once the
problem no longer fits into GPU memory. In this case it is outperformed
by the chunking methods.
 
In $A\times P$ multiplication, $P$ fits into HBM. The Chunk16 method
copies the whole $P$ matrix into HBM and streams through the partitions of $A$ and
$AP$. It obtains optimal data movement cost. As a result, its GFLOP/s 
remain constant for each $A\times P$ multiplication as the data
size gets larger. Note that there is a performance drop for Chunk8 for
32GB BigStar and Brick matrices as the $P$ no longer fits into HBM space for
this portion.

In $R\times A$ multiplication, $A$ is the largest size. Our algorithm
tries to fit $R$ and $RA$ into HBM first. Chunk16 achieves optimal copy
cost for all instances of Elasticity. The optimal data movement is not
satisfied for Laplace, BigStar and Brick matrices for the $A$ with sizes of
16, 16 and 32 GBs, as none of the matrices fits into bigger portion
of the HBM. Still, Chunk16 method achieves 3.10$\times$, 5.36$\times$ and
13.27$\times$ speedups w.r.t. host pinned memory, in the Brick, BigStar and
Elasticity, problems respectively.

GPU architectures can support double and triple buffering, {\em i.e.},
overlapping data movement and computation. When these techniques are
used, the available memory is required to be partitioned into multiple
discrete segments. While some data is copied to one of the segments, the computation
is performed on a separate segment. The Kokkos runtime is currently adding
support for this feature. We do not have implementation using these strategies
but plan to provide an additional analysis using these capabilities in the
near future. However, the performance of Chunk8 on $A \times P$, on most of
$R\times A$ multiplication signals high potential performance gains. 

In conclusion, memory systems in GPU architectures differ
significantly from \knl{}s on latency-based overhead. There is big room for
improvement, and this is explored by the proposed chunking method. We
achieve higher performances when we can fit $B$ or $A$ and $C$ into GPU
memory, as the data movement cost is minimized. The performance drops
when all matrices have to be partitioned. Yet, we obtain significant
speedups w.r.t. UVM and host pinned memory alternatives.


\myspace{-0.75ex}
\section{Conclusion}
\myspace{-0.5ex}

\noindent
Multi-level memory systems have the potential to offer a balance between bandwidth,
latency, capacity, cost and power consumption. Already present in a 
number of supercomputers, multiple-levels of memory are proving to be
a complex target for algorithms which require performance but must operate over
data structure sizes that exceed the capacity of the highest performing pool of memory
in each compute node. 

In this paper we have evaluated methods for performing sparse matrix multiplication on
two complex multi-level memory-enabled architectures -- Intel's Knights Landing
processor and NVIDIA's Pascal GPU. We evaluated 
the performance when placing individual matrices in different memory pools, the use
of hardware-based caching modes, and the chunking of matrices into blocks that
can be copied to/from the most performant memory in a node for
computation. For KNL, the use of hardware cache very often provides performance which
is consistent with an HBM-only execution demonstrating the relatively low overhead the
hardware provides in this mode. GPUs show a different trend, with chunking 
methods being essential once data structures exceed the capacity of the HBM memory
resource. 
As we look to the future, the arrival of yet more complex memory subsystems, particularly
the potential use of non-volatile memory, will make the computing architecture
landscape even more challenging for performant algorithms. {\em Our results suggest that the
design of the multi-level-memory algorithms are crucial for memory systems that significantly 
differ in both latency and bandwidth related metrics. On the other hand, a carefully designed cache-friendly 
standard algorithm can reduce the need of such methods for the architectures that have memory sub-sytems with similar latencies.}
In the future, we would like to extend this study to simulate on SST toolkit to simulate 
on memory subsystems with variable latency and bandwidth overheads.

\small

\noindent {\bf Acknowledgements:} 
We thank the ASC Advanced Architectures test-bed team at Sandia National
Laboratories for supplying and supporting the systems used in this
paper. Additionally, we thank Jonathan Berry and Cynthia Phillips for
helpful discussions during the development of this work.
Sandia National Laboratories is a multimission laboratory managed and
operated by National Technology and Engineering Solutions of Sandia,
LLC., a wholly owned subsidiary of Honeywell International, Inc., for
the U.S. Department of Energy's National Nuclear Security Administration
under contract DE-NA-0003525. 
The research presented in this paper was
funded through the Laboratory Directed Research and Development (LDRD)
program at Sandia National Laboratories, in the context of the
Multi-Level Memory Algorithmics for Large, Sparse Problems Project.
\normalfont




\bibliographystyle{IEEEtranS}

\bibliography{spgemm}

\balance

\end{document}